\documentclass[lettersize,journal]{IEEEtran}
\usepackage{epstopdf}
\usepackage[draft]{hyperref} % optional
\usepackage{setspace}
\usepackage{amsmath}
\usepackage{amssymb}
\usepackage{amsthm}
\usepackage{float}
\usepackage{algorithm}
\usepackage{algorithmicx}
\usepackage{graphicx}
\usepackage{caption}
\usepackage{subcaption}
\usepackage{algpseudocode}
\usepackage{amsfonts}
\usepackage{color}
\usepackage{graphicx}
\usepackage{cite}
\usepackage{mathtools}
\usepackage{stmaryrd}
\usepackage[mathscr]{euscript}
\usepackage{mathrsfs}

\usepackage{soul}

\usepackage{amsthm}

\newtheorem{remark}{Remark} % Defines a numbered remark environment

\newtheorem{theorem}{Theorem}

\newtheorem{cor}{Corollary}

\begin{document}
	\author{Kaluguri Yashaswini*, Anshu Arora*~\IEEEmembership{Student Member,~IEEE} and   Satish~Mulleti~\IEEEmembership{Member,~IEEE}
		\thanks{$^*$ Equal contribution}
		\thanks{The authors are with the Department of Electrical Engineering, Indian Institute of Technology (IIT) Bombay, Mumbai, India.
			Emails: yashkaluguri2005@gmail.com, anshuarora2604@gmail.com,
			mulleti.satish@gmail.com}}

	\title{Adaptive Non-Uniform Sampling of Bandlimited Signals via Algorithm–Encoder Co-Design}
	\markboth{}
	{Shell \MakeLowercase{\textit{Satish Mulleti}}: Adaptive Non-Uniform Sampling}
	\maketitle
	
	\begin{abstract}
		We propose an adaptive non-uniform sampling framework for bandlimited signals based on an algorithm--encoder co-design perspective. By revisiting the convergence analysis of iterative reconstruction algorithms for non-uniform measurements, we derive a local, energy-based sufficient condition that governs reconstruction behavior as a function of the signal and derivative energies within each sampling interval. Unlike classical approaches that impose a global Nyquist-type bound on the inter-sample spacing, the proposed condition permits large gaps in slowly varying regions while enforcing denser sampling only where the signal exhibits rapid temporal variation. Building on this theoretical insight, we design a variable-bias, variable-threshold integrate-and-fire time encoding machine (VBT-IF-TEM) whose firing mechanism is explicitly shaped to enforce the derived local convergence condition. To ensure robustness, a shifted-signal formulation is introduced to suppress excessive firing in regions where the magnitude of the signal amplitude is close to zero or the local signal energy approaches zero. Using the proposed encoder, an analog signal is discretely represented by time encodings and signal averages, enabling perfect reconstruction via a standard iterative algorithm even when the local sampling rate falls below the Nyquist rate. Simulation results on synthetic signals and experiments on ultrasonic guided-wave and ECG signals demonstrate that the proposed framework achieves substantial reductions in sampling density compared to uniform sampling and conventional IF-TEMs, while maintaining accurate reconstruction. The results further highlight a controllable tradeoff between sampling density, reconstruction accuracy, and convergence behavior, which can be navigated through adaptive parameter selection.
	\end{abstract}

	\begin{IEEEkeywords}
		Adaptive non-uniform sampling, time encoding machines, integrate-and-fire models,
		iterative reconstruction, bandlimited signals, event-based sampling,
		algorithm--encoder co-design.
	\end{IEEEkeywords}
	\IEEEpeerreviewmaketitle
\section{Introduction}
Sampling of bandlimited functions is a classical topic with a vast literature. In its simplest form, the Shannon--Nyquist theorem states that a function $f \in PW_{\Omega_0}$ \footnote{We denote by $PW_{\Omega_0}$ the Paley--Wiener space of all functions $f \in L^2(\mathbb R)$ whose Fourier transform is supported in $[-\Omega_0,\Omega_0]$. When needed, we further restrict to functions satisfying a uniform amplitude bound $|f(t)| \le c$, and we write
\[
PW_{\Omega_0,c} \triangleq \{ f \in PW_{\Omega_0} : |f(t)| \le c \}.
\]} can be recovered from its uniform samples $\{f(nT_s)\}_{n\in\mathbb Z}$ provided $T_s \le \pi/\Omega_0$ \cite{nyquist,shannon}. This condition ties the sampling rate to the \emph{maximal} frequency content of $f$ and is therefore governed by the worst-case temporal variation.

In many situations, however, bandlimited functions exhibit highly nonuniform temporal behavior: they vary rapidly only on short intervals and remain slowly varying or nearly flat elsewhere. In such cases, uniform sampling at the Nyquist rate introduces substantial redundancy. This motivates the study of non-uniform sampling schemes, where the sampling set is an arbitrary discrete subset $\Lambda = \{t_n\}_{n\in\mathbb Z} \subset \mathbb R$ and the available data are the samples $\{f(t_n)\}$ or related measurements. Classical density theorems of Landau show that, for functions bandlimited to $[-\Omega_0,\Omega_0]$, uniqueness of reconstruction from $\{f(t_n)\}$ holds provided the lower Beurling density
\[
D^-(\Lambda) \triangleq \liminf_{T\to\infty} \inf_{t\in\mathbb R} \frac{|\Lambda \cap [t,t+T]|}{T}
\]
satisfies $D^-(\Lambda) \ge \Omega_0/\pi$ \cite{landau}. While such conditions characterize sets of uniqueness, they do not in themselves provide stable reconstruction algorithms.

A different line of work investigates sufficient conditions ensuring the convergence of concrete reconstruction procedures from non-uniform samples. Many iterative algorithms for reconstructing bandlimited functions from samples taken at locations $\Lambda = \{t_n\}$ are known to converge under a \emph{global spacing constraint} of the form
\begin{align}
T_n \triangleq t_{n+1}-t_n \le T_{\max} < \frac{\pi}{\Omega_0} \qquad \text{for all } n,
\label{eq:Tn_Tnyq_cond}
\end{align}
see, for example, \cite{feichtinger1995efficient,marvasti1993nonuniform,yao_nonuniform,yen_nonuniform,lazar2004perfect}. Thus, even in the non-uniform setting, many reconstruction schemes effectively enforce a Nyquist-type upper bound on all local gaps, and therefore cannot exploit the presence of long intervals on which $f$ varies slowly.

In parallel, there has been considerable interest in \emph{signal-dependent} sampling mechanisms, in which sampling locations are determined online from the evolving function. Examples include level-crossing schemes and various time encoding machines (TEMs), where samples are generated when the function or a transformed version of it crosses prescribed thresholds; see, for instance, \cite{diff_and_fire,lazar2004time,lazar2004perfect,adam2020encoding,gontier2014sampling,differentTEMS2,differentTEMS3,florescu2022time,thao2020time,kamath2023multichannel,fu2024time,florescu2025generalized,bernardo2025symbol}. In many such constructions, the sampling density is driven primarily by signal amplitude, and existing reconstruction guarantees again rely on a uniform upper bound on the inter-sample spacing, typically of Nyquist type. More recent variants introduce variable bias or variable thresholds in integrate-and-fire TEMs in order to reduce oversampling \cite{AdaptiveBiasIFTEM,hila2022time,naaman2025time}, yet the convergence analyses still depend on global spacing assumptions. Other adaptive non-uniform sampling frameworks, including ``time-stampless'' schemes \cite{feizi2012time,feizi2014backward} and related approaches \cite{dar2019high,pandey2024cardinality,malmirchegini2015non}, are typically analyzed in terms of rate--distortion or energy-efficiency criteria for stochastic or deterministic models, rather than via operator-theoretic convergence in Paley--Wiener spaces.

The present work is motivated by the following question: \emph{Can one design, for bandlimited functions, a sampling scheme in which the local sampling density follows the local variation of the function, while still admitting a provably convergent reconstruction algorithm that does not impose a global Nyquist-type bound on all sample gaps?} Our approach is to separate the problem into two parts. First, we revisit the convergence analysis of a standard iterative reconstruction operator from non-uniform measurements and derive a new \emph{local, energy-based} sufficient condition for convergence. This condition depends on the energies of the function and its derivative on each sampling interval and allows the inter-sample spacing to exceed $\pi/\Omega_0$ on intervals where the function varies slowly. Second, we construct a nonlinear encoding map that, given a bandlimited function $f$, produces a sampling set $\Lambda(f)$ and associated local averages in such a way that the resulting sampling intervals automatically satisfy the required local condition.

More precisely, we consider a reconstruction operator $\mathcal{A}_\Lambda$ associated with a sampling set $\Lambda = \{t_n\}$ and suitable local averages $y_n$ of $f$ over the intervals $[t_n,t_{n+1}]$. For a fixed $\Lambda$ and fixed data $\{y_n\}$, we show that if the intervals $[t_n,t_{n+1}]$ satisfy an inequality of the form
\[
T_n^2 D_n(f) \le \alpha \pi^2 E_n(f) \qquad (0<\alpha<1),
\]
where $E_n(f)$ and $D_n(f)$ denote the $L^2$-energies of $f$ and its derivative $f'$ on $[t_n,t_{n+1}]$, then the operator norm of $\mathcal{I}-\mathcal{A}_\Lambda$ on the Paley--Wiener space is bounded by $\alpha$. In particular, the fixed-point iteration
\[
f_{k+1} = f_k + \mathcal{A}_\Lambda(f - f_k)
\]
converges in $L^2$ to $f$ for any initial $f_0$. This provides a purely local, signal-dependent sufficient condition for convergence that reduces to the classical global spacing condition when the right-hand side is controlled by a uniform bound on $T_n$.

We then construct, for each $f$ in the bandlimited class under consideration, a sampling set $\Lambda(f)$ and local averages $y_n(f)$ using a variable-bias, variable-threshold integrate-and-fire time encoding mechanism of the type studied in \cite{AdaptiveBiasIFTEM,hila2022time,naaman2025time}. The resulting encoder is a nonlinear map which observes $f$ once and outputs a collection of intervals $[t_n,t_{n+1}]$ and corresponding averages $y_n = \int_{t_n}^{t_{n+1}} f(t)\,dt$. We show that, with appropriate choices of the bias and threshold functions (including a shifted-signal formulation to avoid excessive firing when $f$ is small), the sampling intervals produced by this encoder satisfy the local energy condition above. Consequently, for each $f$ the associated reconstruction operator $\mathcal{A}_{\Lambda(f)}$ has $\|\mathcal{I} - \mathcal{A}_{\Lambda(f)}\| < 1$, and the fixed-point iteration based on the fixed data $\{t_n,y_n\}$ converges to $f$.

The main contributions of this paper can be summarized as follows.
\begin{itemize}
  \item We derive a local, energy-based sufficient condition for the convergence of an iterative reconstruction operator from non-uniform measurements of a bandlimited function, replacing the usual global Nyquist-type bound on the inter-sample spacing by a condition expressed in terms of the energies of $f$ and $f'$ on each sampling interval.
  \item We introduce a nonlinear encoding scheme which, given $f$, produces a sampling set $\Lambda(f)$ and local averages $y_n(f)$ via a variable-bias, variable-threshold integrate-and-fire mechanism, and we prove that the resulting sampling intervals satisfy the required local condition.
  \item Combining these two ingredients, we obtain, for each $f$ in a Paley--Wiener space, a non-uniform sampling representation in terms of pairs $(t_n,y_n)$ from which $f$ can be recovered by a fixed iterative algorithm, even though some local gaps $T_n$ may exceed the Nyquist interval on slowly varying regions.
  \item Finally, we illustrate the behavior of the proposed scheme on synthetic and application-inspired examples, highlighting the appearance of long sampling intervals in low-variation regions and demonstrating the convergence of the reconstruction algorithm in practice. Examples include signals motivated by biomedical and ultrasonic-guided-wave applications \cite{Liang1999,Beatty2005}.
\end{itemize}

\section{Adaptive Non-Uniform Sampling Approach}
\label{sec:Adaptive-NUS}

In this section, we present a non-uniform sampling framework in which a bandlimited signal is represented by a set of sampling intervals and corresponding local averages. The key idea is
to generate sampling intervals that are short in regions where the signal varies rapidly and long in regions where the signal varies slowly, while preserving convergence of a standard iterative reconstruction algorithm.

The section is organized as follows. In Subsection~\ref{ssec:algorithm} we review an iterative reconstruction scheme based on local averages and revisit its convergence analysis. This leads to a local, energy-based sufficient condition that replaces the classical global Nyquist spacing constraint \eqref{eq:Tn_Tnyq_cond}. In Subsection~\ref{ssec:VBT-IFTEM} we introduce a variable-bias, variable-threshold integrate-and-fire mechanism that produces sampling intervals satisfying this local condition for each fixed signal. In Subsection~\ref{ssec:biasedVB-IFTEM} we refine the construction to avoid excessive sampling in low-energy regimes. Finally, in Subsection~\ref{ssec:Proposed_AdaptiveNUS} we discuss the proposed adaptive non-uniform sampling.

\subsection{Iterative Reconstruction and Convergence Analysis}
\label{ssec:algorithm}

We first recall an iterative reconstruction procedure based on local averages and examine its convergence properties for a fixed sampling set. Let $f \in PW_{\Omega_0,c}$ and let $\Lambda = \{t_n\}_{n\in\mathbb Z}$ be a sampling set with associated intervals $I_n = [t_n,t_{n+1}]$. Suppose that for each $n$ we are given the local average
\begin{equation}
    y_n = \int_{t_n}^{t_{n+1}} f(t)\,dt.
    \label{eq:signal_averages}
\end{equation}
Given $\Lambda$ and the data $\{y_n\}$, we define a reconstruction operator $\mathcal A$ on $PW_{\Omega_0}$ by
\begin{equation}
   (\mathcal A g)(t) = \sum_{n \in \mathbb{Z}} y_n\, g(t - s_n),
   \label{eq:operator_def}
\end{equation}
where $g(t) = \frac{\sin (\Omega_0 t)}{\pi t}$ is the sinc kernel and
$s_n = \frac{t_{n+1}+t_n}{2}$ is the midpoint of $I_n$. The sums are understood in the $L^2$-sense; for bandlimited $f$ supported in a finite time interval, only finitely many terms contribute.

Using the fixed operator $\mathcal A$ and fixed data $\{y_n\}$ obtained from a \emph{single} encoding of $f$, we consider the iterative scheme
\begin{equation}
    f_{l+1} = f_l + \mathcal{A}(f - f_l),
    \label{eq:reconstruction}
\end{equation}
with initial condition $f_0 = \mathcal A f$. No resampling of the iterates $f_l$ is performed; the sampling set $\Lambda$ and the averages $\{y_n\}$ remain fixed throughout the iteration.

If
\begin{equation}
    \|\mathcal I - \mathcal A\| \le \alpha < 1,
\end{equation}
where $\mathcal I$ denotes the identity operator on $PW_{\Omega_0}$, then it follows that
\begin{equation}
    \| f - f_l \| \le \alpha^{l+1} \| f \|,
\end{equation}
so that $f_l \to f$ in $L^2$ as $l\to\infty$. We therefore seek conditions on $\Lambda$ ensuring that $\|\mathcal I - \mathcal A\| < 1$.

The proof in \cite{lazar2004perfect} proceeds by estimating the norm of $(\mathcal I - \mathcal A)$ in terms of local deviations of $f$ from its midpoint samples. A key step uses Wirtinger's inequality (cf.\ Appendix~B in \cite{lazar2004perfect}) applied to the partial norms
\[
\int_{t_n}^{t_{n+1}} |f(t) - f(s_n)|^2\,dt.
\]
Specifically,
\begin{equation}
    \int_{t_n}^{t_{n+1}} |f(t) - f(s_n)|^2\,dt
    \leq \frac{T_{n}^2}{\pi^2}
    \int_{t_n}^{t_{n+1}} |f'(t)|^2\,dt,
    \label{eq:proof_step}
\end{equation}
where $T_n = t_{n+1}-t_n$. Summing over $n$ yields
\begin{align}
    \sum_{n \in \mathbb{Z}} & \int_{t_n}^{t_{n+1}} |f(t) - f(s_n)|^2\,dt
    = \|\mathcal{I}f - \mathcal{A}^* f\|^2 \notag\\
    &\leq \sum_{n\in\mathbb Z} \frac{T_{n}^2}{\pi^2}
    \int_{t_n}^{t_{n+1}} |f'(t)|^2\,dt,
    \label{eq:proof_step2}
\end{align}
where $\mathcal A^*$ denotes the adjoint of $\mathcal A$ on $L^2(\mathbb R)$. In the classical analysis one then bounds $T_n$ by a global constant $T_{\max}$ and applies Bernstein's inequality $\|f'\|^2 \le \Omega_0^2 \|f\|^2$, which leads to
\begin{equation}
    \|\mathcal I f - \mathcal A f\|^2 \le \left(\frac{T_{\max}\Omega_0}{\pi}\right)^2 \|f\|^2,
    \label{eq:proof_step3}
\end{equation}
and hence imposes the condition $T_{\max}<\pi/\Omega_0$ for convergence.

To obtain a more flexible, local condition, we introduce for each interval $I_n=[t_n,t_{n+1}]$ the energies
\begin{equation}
    E_n(f) \triangleq \int_{t_n}^{t_{n+1}} |f(u)|^2\,du,\qquad
    D_n(f) \triangleq \int_{t_n}^{t_{n+1}} |f'(u)|^2\,du.
    \label{eq:D_E_def}
\end{equation}
With this notation, the right-hand side of \eqref{eq:proof_step2} can be written as
\[
   \|( \mathcal I - \mathcal A^*)f\|^2
   \le \sum_{n\in\mathbb Z} \frac{T_n^2}{\pi^2} D_n(f).
\]

We can now formulate the convergence result purely in terms of these local quantities.

\begin{theorem}[Local energy-based convergence]\label{thm:local-convergence}
Let $f \in PW_{\Omega_0,c}$ and let $\Lambda = \{t_n\}_{n\in\mathbb Z}$ be a sampling set with associated intervals $I_n=[t_n,t_{n+1}]$. Suppose that there exists $0<\alpha<1$ such that
\begin{equation}\label{eq:local-condition}
   \frac{T_n^2}{\pi^2} D_n(f) \le \alpha^2\,E_n(f) \qquad \text{for all } n\in\mathbb Z.
\end{equation}
Let $\mathcal A$ be defined by \eqref{eq:operator_def} using fixed data $\{t_n,y_n\}$ obtained from $f$ via \eqref{eq:signal_averages}. Then
\begin{equation}
   \|\mathcal I - \mathcal A\| \le \alpha
\end{equation}
on $PW_{\Omega_0,c}$. In particular, for any initial $f_0\in PW_{\Omega_0,c}$, the sequence defined by \eqref{eq:reconstruction} satisfies
\begin{align}
   \|f - f_{l+1}\| &= \|( \mathcal I - \mathcal A)(f - f_l)\| \nonumber\\
                   &\le \alpha\,\|f - f_l\| \nonumber\\
                   &\le \alpha^{\,l+1}\,\|f - f_0\|, \qquad l\ge 0,
\end{align}
and hence converges in $L^2$ to $f$. Choosing $f_0 = \mathcal A f$ yields the bound
\begin{equation}
   \|f - f_l\| \le \alpha^{\,l+1}\,\|f\|.
\end{equation}
\end{theorem}

\begin{proof}
By \eqref{eq:proof_step2} and the definition of $D_n(f)$ we have
\begin{equation}
\|(\mathcal I - \mathcal A^*)f\|^2 \le \sum_{n\in\mathbb Z} \frac{T_n^2}{\pi^2} D_n(f).
\end{equation}
Using the local condition \eqref{eq:local-condition}, namely
\[
\frac{T_n^2}{\pi^2} D_n(f) \le \alpha^2 E_n(f) \qquad \text{for all } n,
\]
we obtain
\begin{equation}
\|(\mathcal I - \mathcal A^*)f\|^2
   \le \alpha^2 \sum_{n\in\mathbb Z} E_n(f)
   = \alpha^2 \|f\|^2.
\end{equation}
Taking square roots yields
\begin{equation}
   \|(\mathcal I - \mathcal A^*)f\| \le \alpha \|f\|
   \qquad \text{for all } f\in PW_{\Omega_0,c}.
\end{equation}
Hence $\|\mathcal I - \mathcal A^*\|\le \alpha$ on $PW_{\Omega_0,c}$. Since $\|\mathcal A\| = \|\mathcal A^*\|$ on $L^2(\mathbb R)$, it follows that $\|\mathcal I - \mathcal A\|\le \alpha$ as well. Therefore
\[
\|f_{l+1}-f\| = \|(\mathcal I - \mathcal A)(f_l - f)\|
              \le \alpha \|f_l - f\|
\]
for all $l\ge 0$, so by induction
\[
\|f - f_l\| \le \alpha^{\,l+1}\,\|f - f_0\|.
\]
This shows that the iteration \eqref{eq:reconstruction} converges in $L^2$ to $f$ for any $f_0\in PW_{\Omega_0,c}$. In particular, choosing $f_0 = \mathcal A f$ gives
\[
\|f - f_l\| \le \alpha^{\,l+1}\,\|f\|.
\]
\end{proof}

A simple sufficient condition for \eqref{eq:local-condition} can be expressed directly in terms of the ratio $E_n(f)/D_n(f)$.

\begin{cor}\label{cor:local-gap}
Let $f$ and $\Lambda$ be as in Theorem~\ref{thm:local-convergence}. If the sampling intervals satisfy
\begin{equation}
   T_n < \pi \sqrt{\frac{E_n(f)}{D_n(f)}} \qquad \text{for all } n\in\mathbb Z,
   \label{eq:the_inequality2}
\end{equation}
then the hypothesis \eqref{eq:local-condition} holds with some $0<\alpha<1$, and hence the iteration \eqref{eq:reconstruction} converges in $L^2$ to $f$.
\end{cor}

Condition \eqref{eq:the_inequality2} reveals that larger inter-sample gaps are permissible on intervals where the derivative energy $D_n(f)$ is small relative to the signal energy $E_n(f)$, while smaller gaps are required where $f$ varies rapidly. In particular, the local spacing $T_n$ may exceed the Nyquist interval in slowly varying regions without compromising convergence, provided the neighboring intervals adapt accordingly.

In the following subsections, we construct, for each $f\in PW_{\Omega_0,c}$, a sampling set $\Lambda(f)$ generated by an integrate-and-fire mechanism such that \eqref{eq:the_inequality2} holds, thereby enabling locally adaptive non-uniform sampling with perfect reconstruction\footnote{Throughout the paper, perfect reconstruction implies exact recovery in the limit of iterations.} guarantees.

\subsection{Variable-Bias-Threshold IF-TEM}
\label{ssec:VBT-IFTEM}
\begin{figure}[!t]
    \centering
    \includegraphics[width=0.9\linewidth]{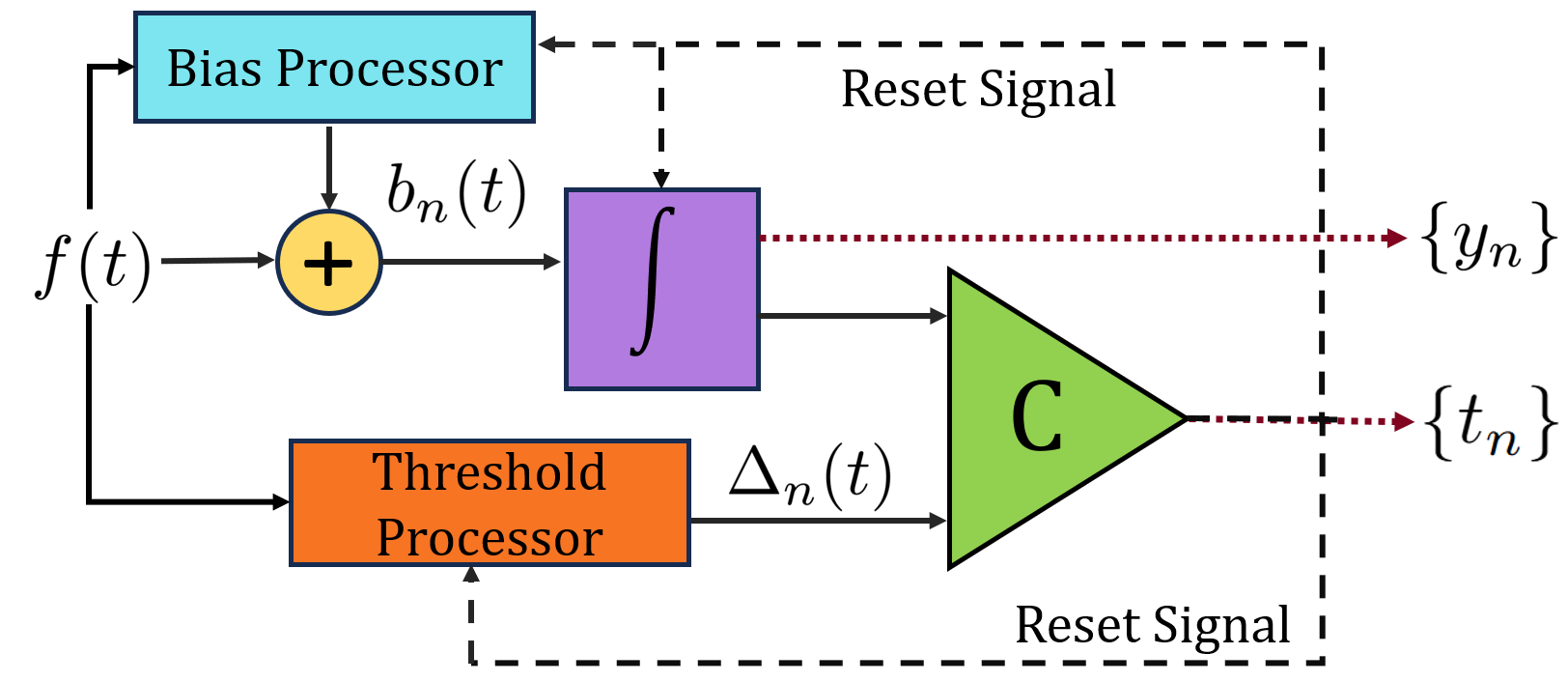}
    \caption{Schematic of the proposed variable-bias, variable-threshold integrate-and-fire encoder.}
    \label{fig:general-TEM}
\end{figure}

We now describe a variable-bias, variable-threshold integrate-and-fire mechanism that produces sampling intervals satisfying the local condition \eqref{eq:the_inequality2}. The generic structure is sketched in Fig.~\ref{fig:general-TEM}. After each firing at time $t_n$, the bias and threshold are updated to functions $b_n(t)$ and $\Delta_n(t)$, respectively, such that $b_n(t)+f(t)>0$ and $\Delta_n(t)>0$ for all
$t \in [t_n, t_{n+1}]$. Following the firing at $t_n$ and the subsequent integrator reset, the
signal $f(t)+b_n(t)$ is integrated and compared to the threshold $\Delta_n(t)$. The next
firing time $t_{n+1}$ satisfies
\begin{equation}
    \int_{t_n}^{t_{n+1}} \big( f(t) + b_n(t)\big)\,dt
    = \Delta_n(t_{n+1}).
    \label{eq:NewTEM_Firing}
\end{equation}

We first derive an upper bound on the interval length $T_n = t_{n+1}-t_n$ generated by this mechanism. The interval $T_n$ is maximized when the integrand $f(t)+b_n(t)$ attains its minimum on $[t_n,t_{n+1}]$. Let
\[
b_{n,\min} = \min_{t \in [t_n, t_{n+1}]} b_n(t).
\]
Using the bound $-c \le f(t)$ for all $t$, we obtain
\begin{equation}
    T_n \leq \frac{\Delta_n(t_{n+1})}{b_{n,\min} - c}.
    \label{eq:upper_Tn_bound}
\end{equation}
Unlike the conventional IF-TEM upper bound $\Delta/(b-c)$, which is constant across intervals, the bound in \eqref{eq:upper_Tn_bound} varies with $n$. By designing $b_n$ and $\Delta_n$ so that the right-hand side of \eqref{eq:upper_Tn_bound} is bounded by the right-hand side of \eqref{eq:the_inequality2}, we ensure that the sampling intervals satisfy the local convergence condition of Corollary~\ref{cor:local-gap}.

The following result formalizes this design for a first choice of bias and threshold.

\begin{theorem}\label{thm:No_bias}
Let $ f \in PW_{\Omega_0, c}$ and let $\{t_n\}$ be the firing times generated by the IF-TEM in Fig.~\ref{fig:general-TEM} with bias and threshold chosen as
\begin{equation}
    b_n(t) = c + \frac{1}{\pi \sqrt{\alpha\,e_n(t)}}, \qquad
    \Delta_n(t) = \frac{1}{\sqrt{d_n(t)+ \beta\,e_n(t)}},
    \label{eq:bias_threshold}
\end{equation}
where $\alpha \in (0,1)$ and $\beta > 0$ are fixed parameters,
\[
e_n(t) = \int_{t_n}^t |f(u)|^2\,du, \qquad
d_n(t) = \int_{t_n}^t |f'(u)|^2\,du,
\]
and $E_n = e_n(t_{n+1})$, $D_n = d_n(t_{n+1})$. Then the sampling intervals $\{T_n\}$ satisfy
\[
T_n < \pi \sqrt{\frac{E_n}{D_n}} \qquad \text{for all } n\in\mathbb Z,
\]
and consequently the local convergence condition of Corollary~\ref{cor:local-gap} holds.
\end{theorem}

\begin{proof}
Since $e_n(t)$ is monotonically increasing on $[t_n,t_{n+1}]$, the bias $b_n(t)$ is decreasing, and its minimum on $[t_n,t_{n+1}]$ is attained at $t_{n+1}$. Thus
\[
b_{n,\min} = c + \frac{1}{\pi\sqrt{\alpha E_n}}.
\]
Substituting the expressions for $b_n$ and $\Delta_n$ into \eqref{eq:upper_Tn_bound} gives
\[
T_n \le \frac{\Delta_n(t_{n+1})}{b_{n,\min}-c}
    = \pi \sqrt{\frac{\alpha E_n}{D_n + \beta E_n}}
    < \pi \sqrt{\frac{E_n}{D_n}},
\]
where the strict inequality follows from $\beta E_n>0$ and $0<\alpha<1$. This is precisely the inequality \eqref{eq:the_inequality2}, so the conclusion follows from Corollary~\ref{cor:local-gap}.
\end{proof}

For reconstruction, we will again use the local averages of $f$ over the intervals $[t_n,t_{n+1}]$. From \eqref{eq:NewTEM_Firing} we can express these as
\begin{equation}
   y_n = \int_{t_n}^{t_{n+1}}  f(t)\,dt
   = \Delta_n(t_{n+1}) - \int_{t_n}^{t_{n+1}} b_n(t)\,dt.
    \label{eq:NewTEM_sigAverage}
\end{equation}
The pairs $(t_n,y_n)$ thus obtained will serve as the discrete representation of $f$.

\begin{remark}
No explicit upper bound is imposed on the maximum gap between firing instants. In particular, $T_n$ may exceed $T_{\text{Nyq}}$ in regions where $E_n$ is large relative to $D_n$. The admissible magnitude of $T_n$ depends on the local signal variations captured by $E_n$ and $D_n$.
\end{remark}

\begin{remark}
\label{remark:need_for_shift}
To illustrate the dependence of the firing rate on signal variation, consider the single-tone
signal $f(t)=A_0\cos(\Omega_m t)$, where $A_0\leq c$ and $\Omega_m\leq\Omega_0$. For this signal,
\begin{equation}
\begin{split}
e_n(t) &= A_0^2 \frac{t-t_n}{2}+A_0^2\frac{\sin (2\Omega_m t)-\sin (2\Omega_m t_n)}{4\Omega_m},\\
d_n(t) &= A_0^2 \Omega_m^2 \frac{t-t_n}{2}-A_0^2 \Omega_m^2\frac{\sin (2\Omega_m t)-\sin (2\Omega_m t_n)}{4\Omega_m}.
\end{split}
\label{eq:en_d_n_cos}
\end{equation}
We regard the signal as varying rapidly when the product $A_0\Omega_m$ is large. Substituting \eqref{eq:en_d_n_cos} into \eqref{eq:bias_threshold} shows that, for fixed $\Omega_m$, increasing $A_0$ decreases both the bias and the threshold, with the threshold decreasing more rapidly, which leads to smaller $T_n$ and hence a higher firing rate. For fixed $A_0$, increasing $\Omega_m$ again decreases the threshold and increases the firing rate. Thus the proposed firing mechanism adapts to local signal variation.
\end{remark}

Despite these desirable properties, the formulation in Theorem~\ref{thm:No_bias} may result in excessive firing when the signal exhibits negligible variation. In particular, when $|f(t)|$ is very small on an interval, both $b_n(t)$ and $\Delta_n(t)$ become large, with the bias term growing faster than the threshold, which can lead to frequent firings even when $e_n(t)\approx 0$.

To mitigate this behavior, one may consider modifying the bias and threshold to prevent unbounded growth, for example by setting
\begin{equation}
\begin{split}
    b_n(t) &= c + \frac{1}{\pi \sqrt{\alpha e_n(t)+\gamma_1^2}},\\
    \Delta_n(t) &= \frac{1}{\sqrt{d_n(t)+ \beta e_n(t)+\gamma_2^2}},
\end{split}
\label{eq:bias_threshold_constnat}
\end{equation}
with parameters $\gamma_1,\gamma_2>0$. Under this modification, when the signal variation is negligible the bias and threshold approach the finite values $c+1/\gamma_1$ and $1/\gamma_2$, respectively, and the firing interval in such regions is bounded by
\[
T_n \leq \frac{\gamma_1}{\gamma_2(c\gamma_1+1)}.
\]
However, it is not immediate from this construction that the inequality \eqref{eq:the_inequality2} still holds. In the next subsection we propose an alternative remedy based on a shifted signal which preserves the theoretical guarantees.

\subsection{VBT-IF-TEM with Biased Signal}
\label{ssec:biasedVB-IFTEM}

To address the issue of oversampling in low-energy regions while preserving the local convergence condition, we introduce a constant shift $s>c$ and apply the encoder to the shifted signal $\tilde f(t) = f(t)+s$. We define the corresponding energies
\begin{equation} 
\tilde e_n(t) = \int_{t_{n}}^t |\tilde f(u)|^2\,du, \qquad
\tilde d_n(t) = \int_{t_{n}}^t |\tilde f'(u)|^2\,du,
\label{eq:tilde_D_E_def} 
\end{equation}
with $\tilde E_n = \tilde e_n(t_{n+1})$ and $\tilde D_n = \tilde d_n(t_{n+1})$. Since $\tilde f(t)>s-c>0$, we have $\tilde e_n(t)>0$ for all $t>t_n$, so the bias and threshold defined in terms of $\tilde e_n$ and $\tilde d_n$ remain finite on every interval.

\begin{theorem}\label{thm:with_bias}
Let $ f \in PW_{\Omega_0, c}$ and fix $s>c$. Consider the IF-TEM in Fig.~\ref{fig:general-TEM} applied to the shifted signal $\tilde f(t) = f(t)+s$, with bias and threshold chosen as
\begin{equation}
    b_n(t) = c - s + \frac{1}{\pi \sqrt{\alpha \tilde e_n(t)}}, \qquad
    \Delta_n(t) = \frac{1}{\sqrt{\tilde d_n(t)+ \beta \tilde e_n(t)}},
    \label{eq:bias_threshold_updated}
\end{equation}
where $\alpha \in (0,1)$ and $\beta > 0$ are fixed parameters and $\tilde e_n,\tilde d_n$ are defined by \eqref{eq:tilde_D_E_def}. Let $\{t_n\}$ be the resulting firing times. Then the intervals $\{T_n\}$ satisfy
\[
T_n < \pi \sqrt{\frac{\tilde E_n}{\tilde D_n}} \qquad \text{for all } n\in\mathbb Z,
\]
and hence the local convergence condition of Corollary~\ref{cor:local-gap} holds for $\tilde f$.
\end{theorem}

\begin{proof}
As before, $\tilde e_n(t)$ is increasing, hence $b_n(t)$ is decreasing on $[t_n,t_{n+1}]$ with minimum
\[
b_{n,\min} = c - s + \frac{1}{\pi\sqrt{\alpha \tilde E_n}}.
\]
Since $\tilde f(t) > s-c$, the integrand $\tilde f(t)+b_n(t)$ is bounded below by $b_{n,\min}+s-c$, so the analogue of \eqref{eq:upper_Tn_bound} yields
\[
T_n \le \frac{\Delta_n(t_{n+1})}{b_{n,\min} + s - c}.
\]
Substituting \eqref{eq:bias_threshold_updated} gives
\begin{align}
        T_n \leq \frac{\Delta_n(t_{n+1})}{b_{n,\min} +s -c} = \pi \sqrt{\frac{ \alpha \tilde E_n}{\tilde D_n + \beta \tilde E_n}} < \pi \sqrt{\frac{\tilde E_n}{\tilde D_n}}, 
        \label{eq:upper_Tn_bound2}
    \end{align}
which is again of the form \eqref{eq:the_inequality2}. The claim then follows from Corollary~\ref{cor:local-gap}, applied to $\tilde f$.
\end{proof}

\begin{figure}[!t]
    \centering
    \begin{subfigure}[b]{0.48\linewidth}
        \centering
        \includegraphics[width=\linewidth]{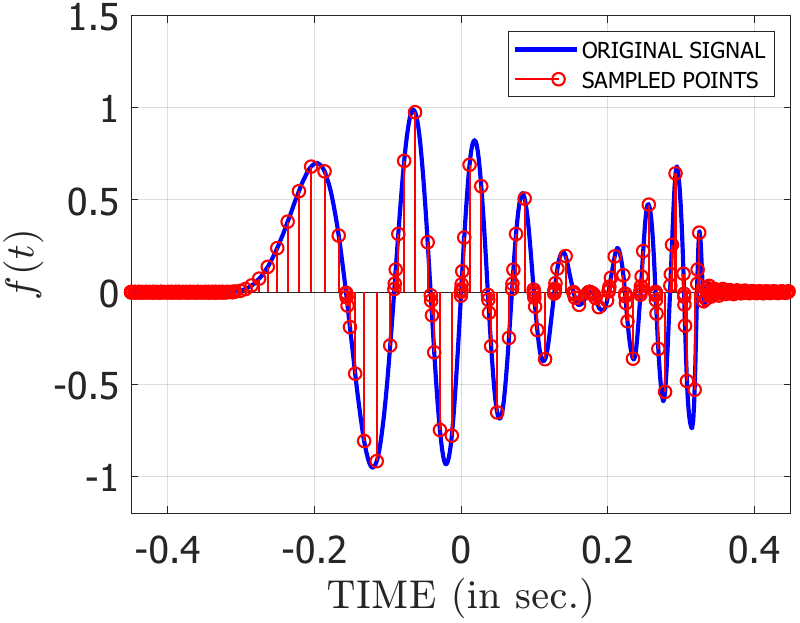}
        \caption{}
        \label{fig:over-sampling}
    \end{subfigure}
    \hfill
    \begin{subfigure}[b]{0.48\linewidth}
        \centering
        \includegraphics[width=\linewidth]{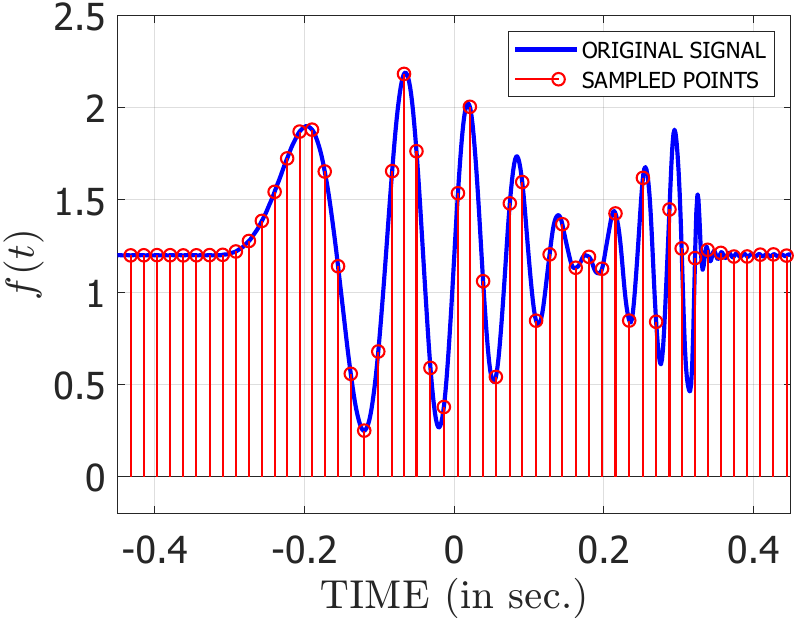}
        \caption{}
        \label{fig:sampling}
    \end{subfigure}
    \caption{A comparison of firings of VBT-IF-TEMs (a) without shift -- 396 firings, and (b) with shift -- 51 firings. }
    \label{fig:sampling-comparison}
\end{figure}

For the shifted signal $\tilde f$, the reconstruction operator $\mathcal A$ built from the averages of $\tilde f$ over the intervals $[t_n,t_{n+1}]$ satisfies $\|\mathcal I - \mathcal A\|<1$, so the iteration \eqref{eq:reconstruction} converges to $\tilde f$. Since $f = \tilde f - s$, recovering $f$ reduces to subtracting the constant offset.
%%%%%%%%%%%%%%%%%%%%%%%%%%%%%%%%%%%%
With the additional signal shift, the proposed sampling scheme retains all the desirable
properties established in Theorem~\ref{thm:No_bias}, while avoiding excessive firing when the
signal amplitude is small. This behavior can be verified by reconsidering the
single-tone example discussed in Remark~\ref{remark:need_for_shift}. To further illustrate the
effect of the signal shift, Fig.~\ref{fig:sampling-comparison} compares the firing patterns of
the VBT-IF-TEM with and without the shift for a representative bandlimited signal. We note that the firings have been reduced from 396 to 51 by adding a bias.

As anticipated from the analysis in Remark~\ref{remark:need_for_shift}, the unshifted formulation
exhibits excessive firing in low-energy regimes, particularly in regions where the magnitude of the signal amplitude is small. In contrast, introducing the shift
controls the growth of the bias term $b_n(t)$, thereby significantly reducing the firing rate in
such regions while preserving dense sampling in regions of high signal variation.

To further elucidate the behavior of the proposed scheme and to quantify the dependence of the
firing rate in response to signal variation, we consider the following two representative scenarios.

\begin{itemize}
    \item \emph{Signal with low variation:} In this case, both the signal and its derivative are
    negligible over the interval $T_n$. Mathematically, this corresponds to $\tilde D_n \to 0$
    and $f(t) \to 0$, which implies $\tilde f(t) \to s$. Substituting into
    \eqref{eq:upper_Tn_bound2}, we obtain
\begin{align}
    T^{\text{Low}}_n \leq \pi \sqrt{\frac{\alpha}{\beta}} = T_{\max}.
    \label{eq:low_variation_bound_Tn}
\end{align}
This bound indicates that when the signal variation is negligible, the firing intervals are
large and can be controlled through appropriate choices of $\alpha$ and $\beta$.
    \item \emph{Signal with large variation:} In this scenario, both the signal amplitude and its
    derivative are significant. Specifically, we consider $\tilde f'(t) \to c\Omega_0$ and
    $f(t)\to c$, which implies $\tilde f(t)\to s+c$. Substituting into
    \eqref{eq:upper_Tn_bound2}, we obtain
\begin{align}
    T^{\text{High}}_n &\leq \pi \sqrt{\frac{\alpha(s+c)^2}{\beta(s+c)^2 + (c\Omega_0)^2}}, \nonumber\\
    &\leq \pi \sqrt{\frac{\alpha}{\beta}}
    \sqrt{\frac{(s+c)^2 \beta}{(c\Omega_0)^2 + (s+c)^2 \beta}}.
    \label{eq:high_variation_bound_Tn}
\end{align}
Here, $T^{\text{High}}_n$ denotes the inter-firing interval in regions of high signal variation.

    The upper bound in the high-variation regime is tighter than that in the low-variation case,
    which is consistent with the desired behavior of increased firing density during rapid
    signal changes. Moreover, the bounds in
    \eqref{eq:low_variation_bound_Tn} and \eqref{eq:high_variation_bound_Tn} can be controlled
    through suitable choices of $\alpha$, $\beta$, and $s$, providing explicit tuning of the
    sampling density across different signal regimes.
\end{itemize}

With the biased VBT-IF-TEM, the local convergence condition is satisfied, and excessive firing is
effectively suppressed in low-energy regimes where the magnitude of the signal amplitude is small. We then formalize an adaptive non-uniform sampling representation based on the resulting time encodings and signal averages.

\subsection{Proposed Adaptive Non-Uniform Sampling}
\label{ssec:Proposed_AdaptiveNUS}

The proposed adaptive non-uniform sampling approach is built upon the VBT-IF-TEM framework discussed in the previous subsections. In its biased form, the TEM produces a sequence of firing times $\{t_n\}$ together with the local signal averages $y_n$ defined in \eqref{eq:NewTEM_sigAverage}. These quantities together constitute a discrete representation of $f$.

In a conventional IF-TEM with constant bias and threshold, the signal averages can be recovered from the time encodings, yielding $\Delta - b(t_{n+1}-t_n)$ from the inter-firing intervals. In contrast, when the bias and threshold vary across firing intervals, recovering the averages during reconstruction requires explicit knowledge of the corresponding bias and threshold functions. To avoid this additional overhead, the proposed scheme computes and stores the signal averages directly during the encoding process, as illustrated in Fig.~\ref{fig:general-TEM}. This leads to the following result.

\begin{theorem}[Adaptive non-uniform sampling]\label{thm:adaptive-NUS}
Let $ f \in PW_{\Omega_0, c}$ and let $\{t_n\}$ denote the firing times generated by the biased VBT-IF-TEM of Theorem~\ref{thm:with_bias}. Let
\begin{equation}
y_n = \int_{t_n}^{t_{n+1}} f(t)\,dt
\end{equation}
be the associated local averages given by \eqref{eq:NewTEM_sigAverage}. Then the collection $\{(t_n,y_n)\}_{n\in\mathbb Z}$ provides a discrete representation of $f$ with the following property: the reconstruction operator $\mathcal A$ defined by \eqref{eq:operator_def} using these fixed data satisfies $\|\mathcal I - \mathcal A\|<1$ on $PW_{\Omega_0,c}$, and the iteration \eqref{eq:reconstruction} converges in $L^2$ to $f$ for any initial $f_0\in PW_{\Omega_0,c}$.
\end{theorem}

While the VBT-IF-TEM in its basic form produces only time encodings, the proposed adaptive non-uniform sampling framework additionally encodes signal averages; this distinction enables stable reconstruction from the discrete representation $\{(t_n,y_n)\}$ via the iterative algorithm of Section~\ref{ssec:algorithm}. In particular, the sampling density is high only in regions where $f$ exhibits rapid variation, and can be arbitrarily low in slowly varying regions, while the reconstruction algorithm remains convergent for the fixed sampling set associated with each $f$.

\begin{figure}[!t]
    \centering    \includegraphics[width=0.95\linewidth]{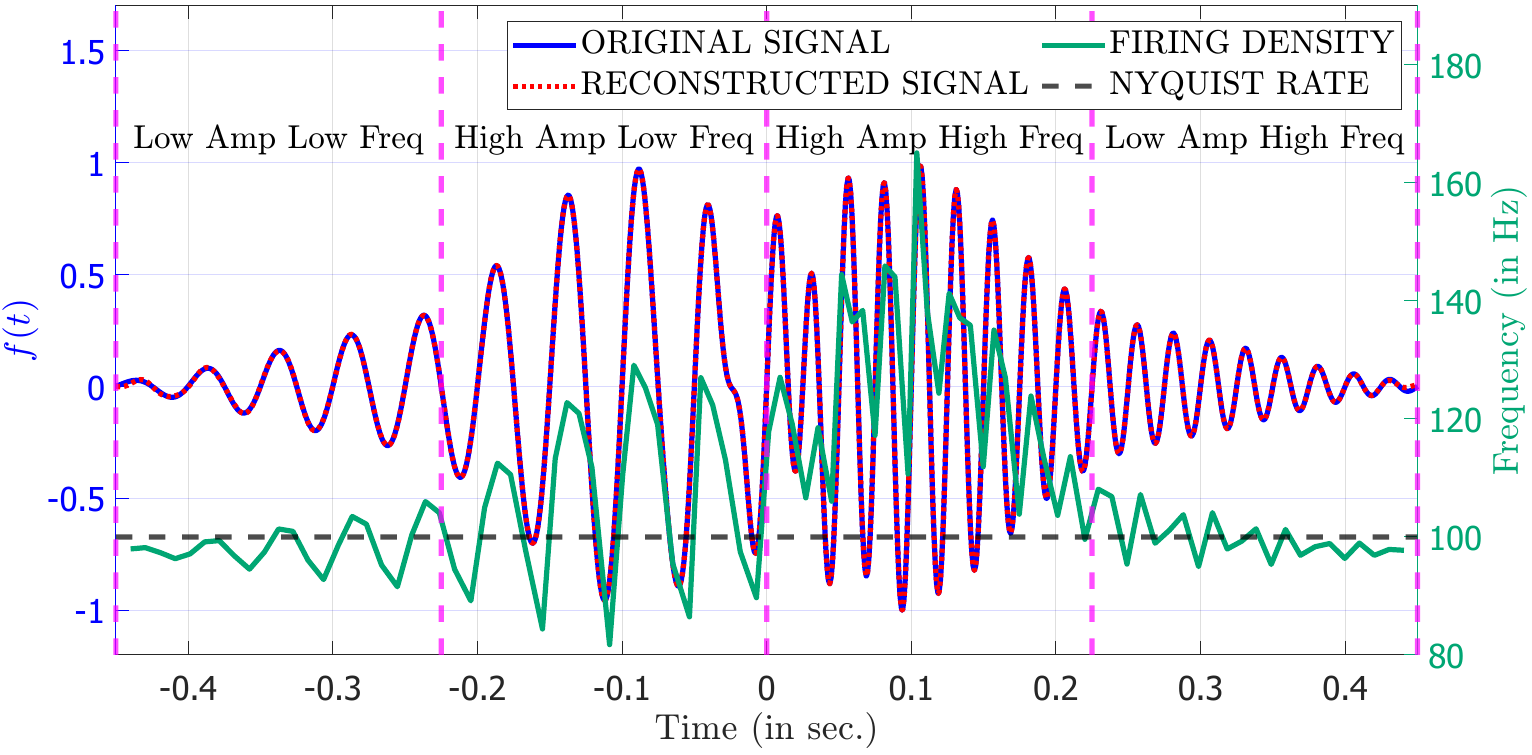}
    \caption{Different firing rates in different regions}
    \label{fig:four_regions_formatted_overlaid}
\end{figure}

To gain further insight into the behavior of the proposed adaptive sampling strategy, we analyze its performance on a representative signal exhibiting four distinct regions of variation, as illustrated in Fig.~\ref{fig:four_regions_formatted_overlaid}. This example allows us to examine how the firing rate, computed as $1/T_n$, adapts to changes in signal amplitude and local frequency content. As expected, the firing rate remains low in regions of low amplitude and low variation, and increases significantly as the signal amplitude and local frequency increase. Notably, in certain regions, the firing rate falls below the Nyquist rate without compromising perfect reconstruction.

In the example of Fig.~\ref{fig:four_regions_formatted_overlaid}, the sampling parameters used are $\alpha=0.7$, $\beta=2450$, $c=1$, and $s=3c$. The proposed method uses $95$ samples, compared to $90$ samples required by uniform sampling at the Nyquist rate, and the resulting reconstruction achieves a normalized mean squared error (NMSE) of $-52.24~\mathrm{dB}$. A comparison with uniform Nyquist sampling and conventional IF-TEM schemes highlights the ability of the biased VBT-IF-TEM to concentrate samples in regions of high signal variation while suppressing unnecessary firings in low-energy regimes.

\subsection{Related Work}
In this subsection, we position the proposed framework within the broader literature on
non-uniform sampling and reconstruction of bandlimited signals, with an emphasis on the theoretical convergence guarantees established in Theorems~\ref{thm:No_bias}, \ref{thm:with_bias}, and \ref{thm:adaptive-NUS}.

\paragraph*{Irregular sampling and reconstruction}
Early work on irregular sampling primarily focused on the \emph{sampling jitter} model, where uniform sampling locations are perturbed by small timing errors
\cite{duffin1952class,wiley1978recovery,maravasti1991}. A seminal contribution by Duffin and Schaeffer~\cite{duffin1952class} established a frame-based reconstruction framework for such sampling patterns. Subsequent works extended these ideas to more general non-uniform sampling sets and iterative reconstruction algorithms
\cite{marvasti1993nonuniform,grochenig1992reconstruction}; see
\cite{feichtinger2021theory} for a comprehensive review.

Several studies have also considered reconstruction from \emph{signal averages} rather than pointwise samples
\cite{grochenig1992reconstruction,sun2002reconstruction,song2012_avgSampling,feichtinger2021theory}, which forms the mathematical basis for IF-TEM reconstruction. In these approaches, convergence of the iterative algorithm is guaranteed under a global Nyquist-type constraint on the maximum inter-sample spacing, as expressed in~\eqref{eq:Tn_Tnyq_cond}. A related line of work investigates accelerating convergence by oversampling, showing that enforcing $T_{\max}<T_{\text{Nyq}}/m$ yields a convergence factor of $\alpha^m$, where $m \in \mathbb{N}$
\cite{grochenig1992reconstruction,marvasti1993nonuniform,li2006remarks}.

\paragraph*{Adaptive and signal-dependent sampling}
Adaptive sampling schemes based on \emph{level crossings} have been widely studied, wherein samples are acquired at time instants when the signal crosses predefined amplitude levels
\cite{logan1977information,mark1981nonuniform,sayiner1996level,boche2012towards,
rzepka2018reconstruction,tsivids2010tutorial}. For a fixed set of levels $\{\Delta_k\}_{k=1}^K$, the signal is represented by time instants $\{t_{k,n}\}$ satisfying $f(t_{k,n})=\Delta_k$. While increasing the number of levels results in denser sampling, the sampling density depends strongly on the level placement within the signal's dynamic range. Although the uniqueness of reconstruction can be ensured when the crossing density exceeds the Nyquist rate \cite{logan1977information,boche2012towards}, explicit conditions ensuring convergence of iterative reconstruction algorithms—particularly guarantees analogous to \eqref{eq:Tn_Tnyq_cond}—remain largely unexplored.

In practice, when a signal remains between two successive levels for extended durations, inter-sample gaps may become arbitrarily large. Whether such sampling patterns satisfy the
requirements for stable reconstruction is generally unclear. Some works attempt to regulate the crossing density by adding auxiliary signals, such as ramps \cite{martinez2016amplitude}, but a systematic link between signal variation and reconstruction guarantees is missing.

\paragraph*{Nonlinear transformations and varying-bandwidth models}
Alternative approaches to non-uniform sampling include nonlinear transformations \cite{clark1985transformation,dvorkind2008nonlinear} and models with time-varying bandwidth
\cite{horiuchi1968sampling,wei2007sampling,rzepka2013recovery,andreolli2023sampling}. For example,
\cite{clark1985transformation} proposes a nonlinear mapping that converts non-uniform samples $\{f(t_n)\}$ into uniform samples of a transformed bandlimited signal. Similarly, time-varying bandwidth models allow local adaptation of sampling density. However, these methods rely on restrictive signal models or transformations, limiting their applicability. In contrast, the proposed framework applies to all bandlimited signals without imposing additional structural assumptions.

\paragraph*{IF-TEM with variable bias}
Recent works on IF-TEMs with variable bias have demonstrated that adapting the bias can reduce oversampling relative to fixed-bias designs, while the firing rate remains primarily proportional to the signal amplitude \cite{omar2024adaptive,VBIFTEM}. In contrast, the present work generalizes the IF-TEM paradigm by jointly adapting both the bias and the threshold as functions of local signal energies. This design explicitly links firing behavior to signal variation and is theoretically grounded in local convergence guarantees for iterative reconstruction.

\paragraph*{Summary}
In summary, while prior work has addressed irregular sampling, adaptive acquisition, and IF-TEM encoding from different perspectives, the proposed framework uniquely integrates a local, energy-based convergence analysis with signal-dependent time encoding. This integration enables adaptive non-uniform sampling with rigorous reconstruction guarantees that do not rely on a global Nyquist-type constraint.

\section{Simulation Results}
\label{sec:simulation_results}

In this section, we compare the proposed adaptive non-uniform sampling method with two baseline
approaches: (i) uniform sampling at or above the Nyquist rate, and (ii) a conventional IF-TEM (C-IF-TEM) with fixed bias and threshold. For all methods, signal reconstruction is performed using signal averages and the iterative reconstruction algorithm described earlier.

In all experiments, we consider bandlimited signals with finite temporal support, such that the signal is negligible outside the observation interval. Performance is evaluated in terms of the number of samples ($\#S$) and the normalized mean-squared error (NMSE), defined as
\begin{align}
    \text{NMSE (in dB)} = 10 \log
    \frac{\|f(t)-\hat{f}(t)\|^2}{\|f(t)\|^2},
    \label{eq:NMSE}
\end{align}
where $\hat{f}(t)$ denotes the reconstructed signal. For reconstruction and NMSE evaluation, both
$f(t)$ and $\hat{f}(t)$ are evaluated on a fine uniform time grid.

We consider two classes of bandlimited signals: a chirp-like signal and a sum-of-sincs (SoS)
signal. These signal classes capture key characteristics of practical signals, including strong
temporal nonuniformity and localized regions of high variation.

\subsection{Chirp Signal}

We first consider a chirp-like bandlimited signal, which is particularly well-suited for demonstrating the benefits of adaptive sampling. Such signals exhibit pronounced \emph{temporal nonuniformity}, consisting of extended intervals of slow variation interspersed
with short bursts of rapid change. This structure allows sparse sampling in low-activity
regions, while the rapidly varying segments dictate the overall bandwidth.

The test signal is constructed as a bandlimited waveform via sinc interpolation, where the
interpolation coefficients are modulated by a slowly varying envelope multiplied by a nonlinear
chirp. This results in localized bursts of high temporal variation within an otherwise smooth
signal. Specifically, the signal is generated as
\begin{align}
f(t)
&= \sum_{m=1}^{M}
c_m \,
\mathrm{sinc}\!\left(
2 F_0
\left(
t - mT_s + \frac{M}{2}T_s
\right)
\right), \label{eq:bl_signal} \\
F_0 &=  100~\text{Hz}, \quad
T_s = \frac{1}{2 F_0}, \quad
M = 130, \nonumber \\
c_m
&= \sin\!\bigl(2\pi \times 0.005\, m\bigr)
\sin\!\left(
2\pi \times 0.081 \,
\frac{m^{2.1}}{2M}
\right), \nonumber \\
t &\in [-T_{\max},\, T_{\max}], \quad
T_{\max} = 0.45~\mathrm{s}. \nonumber
\end{align}
By definition, $\mathrm{sinc}(x)=\sin(\pi x)/(\pi x)$, and the signal amplitude is normalized to satisfies
$|f(t)|\leq c=1$. For this experiment, the parameters of the proposed method are set to
$\alpha=0.5$, $\beta=5600$, and $s=4.2$, while $b=1.3$ and $\Delta=0.0015$ were chosen for the C-IF-TEM. These parameters are selected to yield the lowest NMSE for each method.

\begin{figure*}[!t]
    \centering
    \renewcommand{\arraystretch}{1.2}
    \begin{tabular}{ccc}
        \begin{minipage}{0.3\linewidth}
            \centering
            \includegraphics[width=\linewidth]{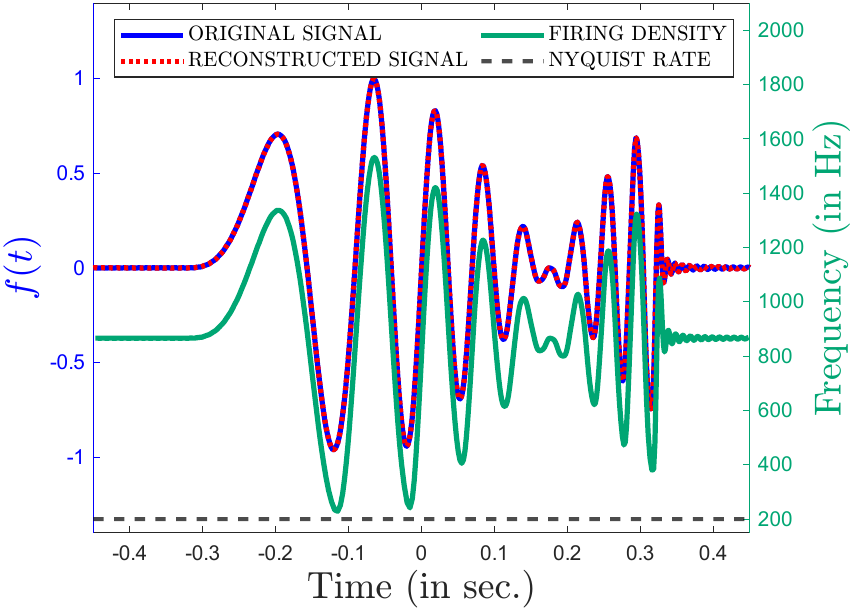}
            \subcaption{C-IF-TEM: $\# S = 796$; NMSE $= -59.96$ dB }
        \end{minipage}
        &
        \begin{minipage}{0.3\linewidth}
            \centering
            \includegraphics[width=\linewidth]{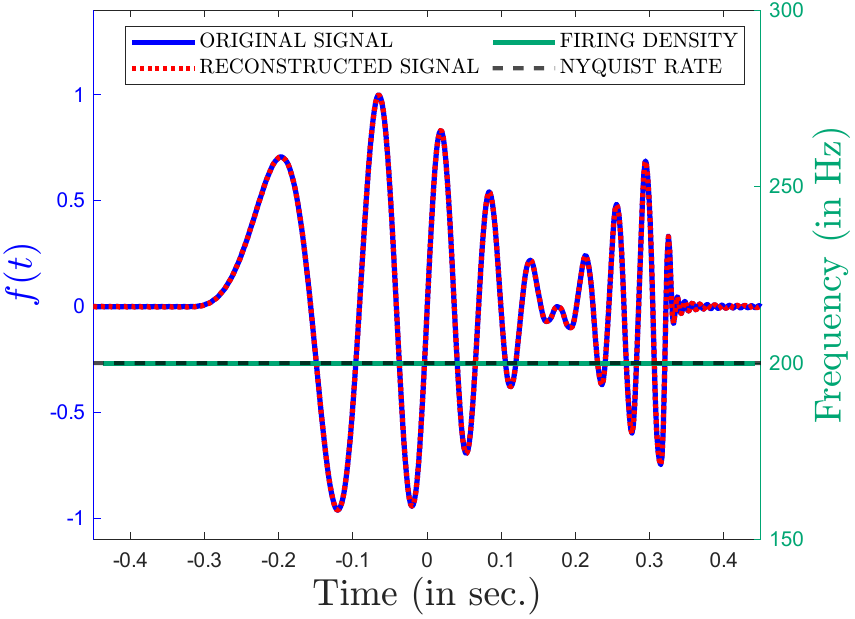}
            \subcaption{Uniform sampling: $\#S = 180$; NMSE $= -62.37$ dB}
        \end{minipage}
        &
                \begin{minipage}{0.3\linewidth}
            \centering
            \includegraphics[width=\linewidth]{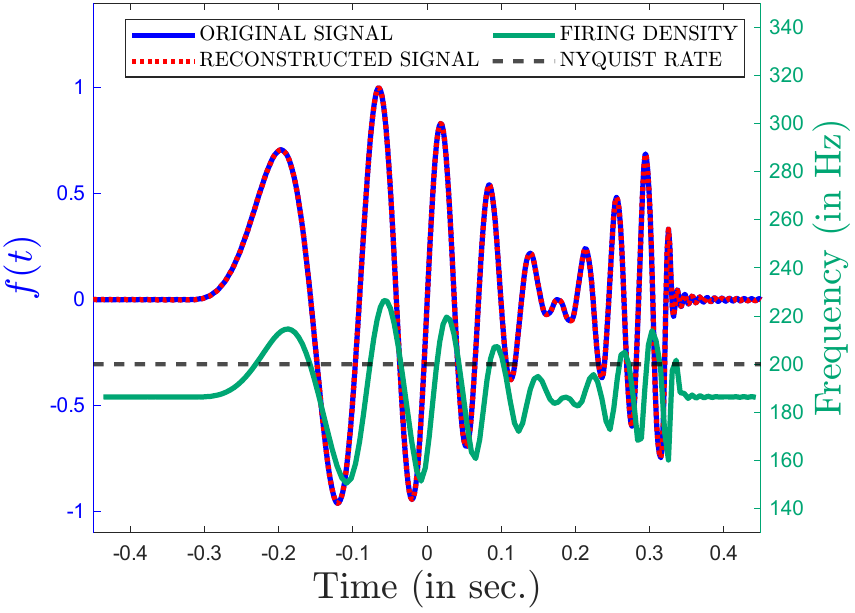}
            \subcaption{Adaptive-NUS: $\#S= 169$; NMSE = $-53.99$ dB}
        \end{minipage} \\
    \end{tabular}
\caption{Comparison of different sampling schemes for a chirp-like signal. The proposed Adaptive-NUS produces fewer samples and achieves sub-Nyquist sampling in low-activity regions.}
    \label{fig:comparison-table_favo}
\end{figure*}

Fig.~\ref{fig:comparison-table_favo} compares the reconstruction results and sampling patterns
obtained using uniform sampling, C-IF-TEM, and the proposed adaptive method. We point out that, in all our experiments, the reconstruction used for uniform sampling is sinc interpolation, and the iterative reconstruction method is used for the other two sampling schemes. Several observations
can be made. First, all methods achieve high reconstruction accuracy, with NMSE values below
$-50$~dB. Second, both uniform sampling and the proposed method require substantially fewer
samples than C-IF-TEM. Third, the sampling density of the proposed method adapts to the local
signal variation and can fall below the Nyquist rate in low-activity regions.

Within the observation interval, the proposed method uses $169$ samples, compared to $180$
samples required by uniform Nyquist sampling. In practice, the proposed method requires storing
both time encodings and signal averages, resulting in $2\times169$ measurements. This modest
increase in measurement count comes with the advantage of eliminating the need for a global clock and the ability to
sample below the Nyquist rate in extended low-activity regions, which can yield significant
savings in applications characterized by sparse signal activity.

\begin{figure*}[!t]
    \centering
    \renewcommand{\arraystretch}{1.2}

    \begin{tabular}{ccc}
        % \textbf{Running IF-TEM (Our Method)} 
        % & \textbf{Conventional IF-TEM} 
        % & \textbf{Uniform Iterative} 
        % \\[0.4em]

        % ---------- Single Row of 3 Images ----------

        \begin{minipage}{0.3\linewidth}
            \centering
            \includegraphics[width=\linewidth]{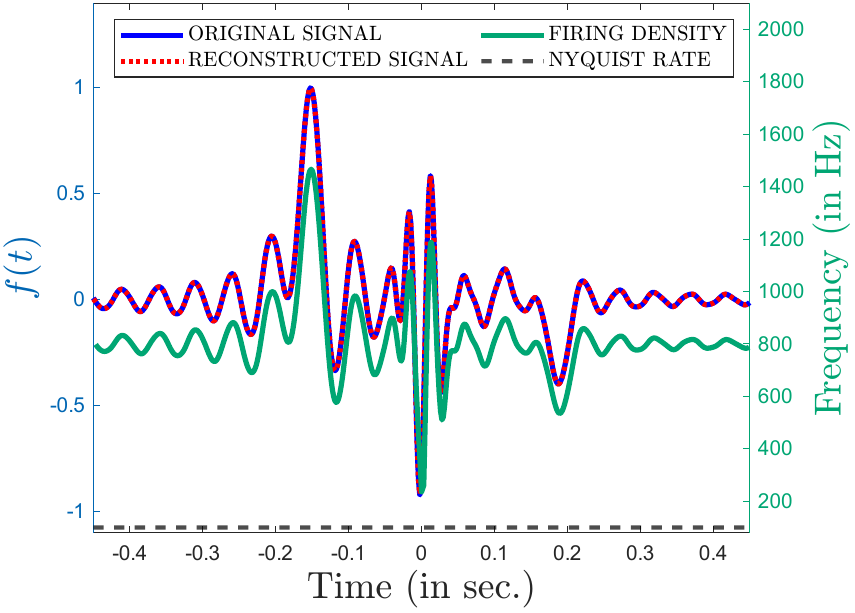}
            \subcaption{C-IF-TEM: $\#S = 729$; NMSE $= -55.82$ dB }
        \end{minipage}
        &
        \begin{minipage}{0.3\linewidth}
            \centering
            \includegraphics[width=\linewidth]{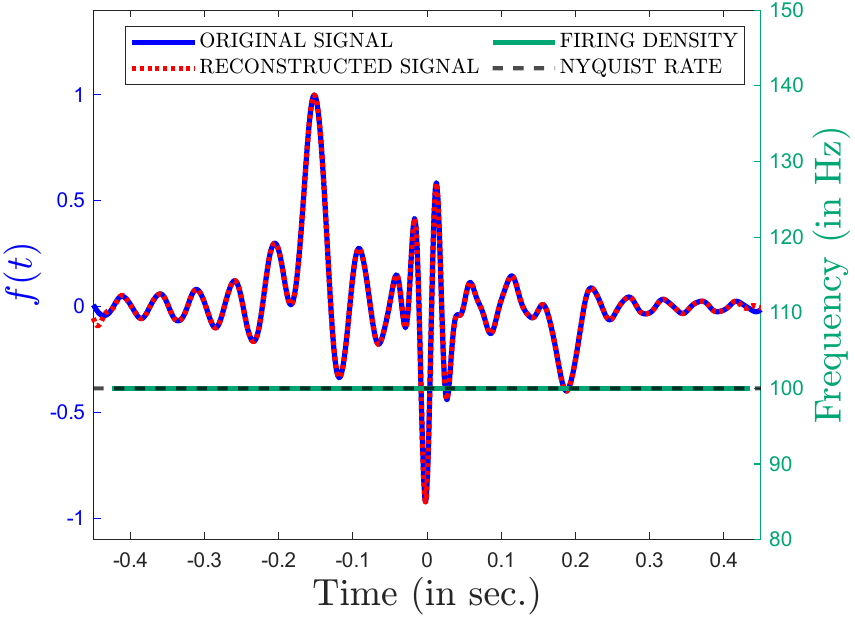}
            \subcaption{Uniform sampling: $\#S = 90$; NMSE = $-52.44$ dB}
        \end{minipage}
        &
                \begin{minipage}{0.3\linewidth}
            \centering
            \includegraphics[width=\linewidth]{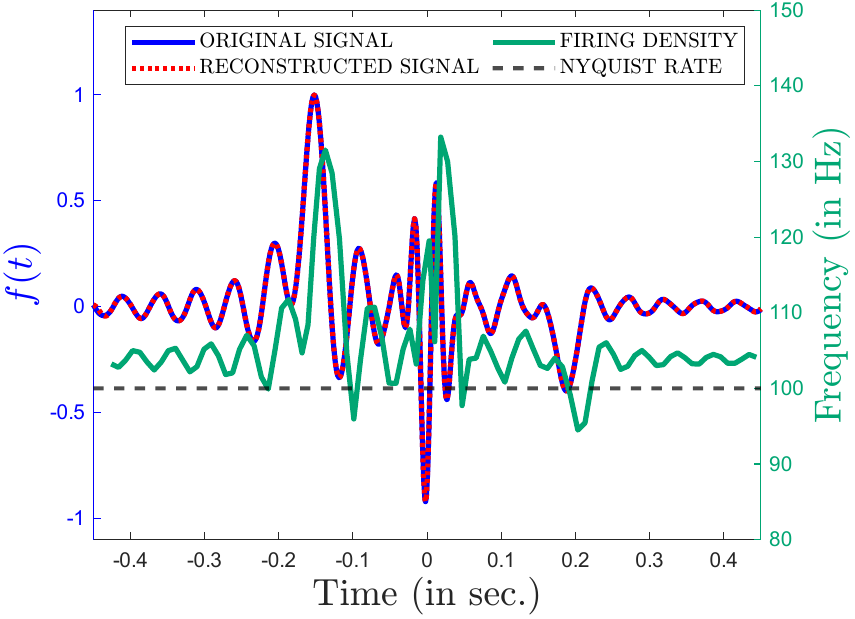}
            \subcaption{Adaptive-NUS: $\#S = 96$; NMSE $= -61.38$ dB}
        \end{minipage}
        
        \\

        % \multicolumn{3}{c}{
        %     \textbf{Comparison of signal reconstruction across three methods}
        % }
    \end{tabular}

    \caption{Comparison of different sampling schemes for a SoS signal with multiple bandwidth regions. The proposed Adaptive-NUS produces fewer samples and achieves sub-Nyquist sampling in low-activity regions.}
    \label{fig:comparison-table_sinc}
\end{figure*}

\subsection{Sum-of-Sincs Signal}
We next evaluate the proposed method using a sum-of-sincs (SoS) signal, which provides a natural
and flexible representation of bandlimited signals. This formulation enables the construction
of signals with multiple regions of distinct bandwidths and varying degrees of temporal
activity.

For the simulations, the signal is generated as a non-overlapping linear combination of sinc
functions with different bandwidths, given by
\begin{align}
    f(t) &= f_1(t) + f_2(t-\tau) + f_2(t+\tau),
    \label{eq:double_sinc_sum}
\end{align}
where
\begin{align}
    f_k(t) &= \sum_{n = -N_k}^{N_k} c_{k, n} \text{sinc}(2F_k(t-nT_k)),~ \text{for}~k = \{1,2\}, \nonumber 
\end{align}
with $T_1 = 0.6 ~\text{ms},~T_2 = 0.4~\text{ms}, ~\tau = 0.15~\text{s},~ N_1 = 50, ~N_2 = 100,~ c_{k,n} \sim \text{Uniform}(-0.5,\,0.5), ~F_1 = 50 \text{~Hz}, ~F_2 = 20 \text{~Hz}$. The resulting signal is normalized to unit amplitude and truncated to the interval $[-0.45,\,0.45]$~s. It consists of three distinct regions: the first and last segments are dominated by slowly varying components with a bandwidth $20$~Hz, while the middle segment is a faster signal with a bandwidth $50$~Hz.

Fig.~\ref{fig:comparison-table_sinc} shows the signal, its reconstructions, and the corresponding firing patterns. Consistent with the chirp experiment, the proposed method's sampling density increases with the local signal's amplitude and frequency content. We use the parameters of the proposed method are set to
$\alpha=0.45$, $\beta=2400$, and $s=3$, while the C-IF-TEM parameters are chosen as
$b=1.2$ and $\Delta=0.0015$.

To obtain statistically meaningful results, we evaluate the methods over $100$ independent SoS signals. The average NMSEs and sampling rates are reported in Table~\ref{tab:sampling_nmse}, and closely match the behavior observed in individual trials.
\begin{table}[t]
\centering
\caption{Comparison of sampling strategies}
\label{tab:sampling_nmse}
\resizebox{0.47\textwidth}{!}{%
\begin{tabular}{lcc}
\hline
\textbf{Sampling Type} & \textbf{Number of Samples} & \textbf{NMSE Error (dB)} \\
\hline
Uniform Sampling      & 90  & $-44.76$ \\
Conventional IF-TEM   & 731 & $-55.70$ \\
Adaptive-NUS            & 96  & $-59.74$ \\
\hline
\end{tabular}}
\end{table}

\begin{figure}[!t]
    \centering
    \includegraphics[width=0.75\linewidth]{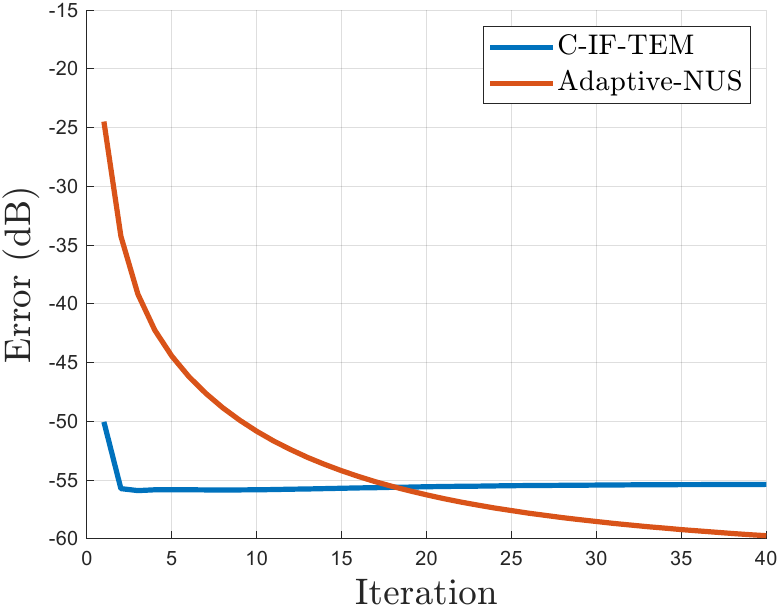}
    \caption{NMSE across 100 SoS signals versus iteration count for the iterative reconstruction algorithm.}
    \label{fig:comparison_iteration}
\end{figure}

We next examine the convergence behavior of the iterative reconstruction algorithm. Since the
proposed method relies on locally adaptive convergence conditions rather than a global bound, a
slower convergence rate is expected compared to C-IF-TEM. Fig.~\ref{fig:comparison_iteration}
shows the average NMSE across $100$ SoS signals as a function of iteration count.

As anticipated, the proposed method requires more iterations to converge. However, this increased iteration count is accompanied by a substantial reduction in the number of samples. Specifically, C-IF-TEM requires approximately $731$ samples on average, whereas the proposed method achieves comparable reconstruction accuracy with only about $96$ samples on average. Increasing the number of iterations beyond the values shown in
Fig.~\ref{fig:comparison_iteration} does not alter the final reconstruction error, confirming that the observed behavior reflects a tradeoff between sampling density and convergence speed rather than a loss of reconstruction fidelity.

\begin{figure}[!t]
    \centering
    \begin{subfigure}[b]{0.48\linewidth}
        \centering
        \includegraphics[width=\linewidth]{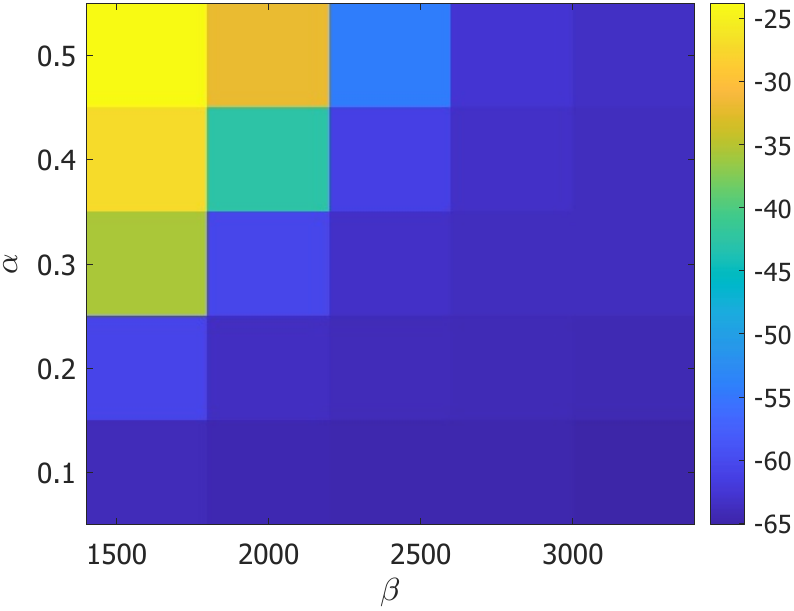}
        \caption{Average NMSE}
        \label{fig:heatmap-error}
    \end{subfigure}
    \hfill
    \begin{subfigure}[b]{0.48\linewidth}
        \centering
        \includegraphics[width=\linewidth]{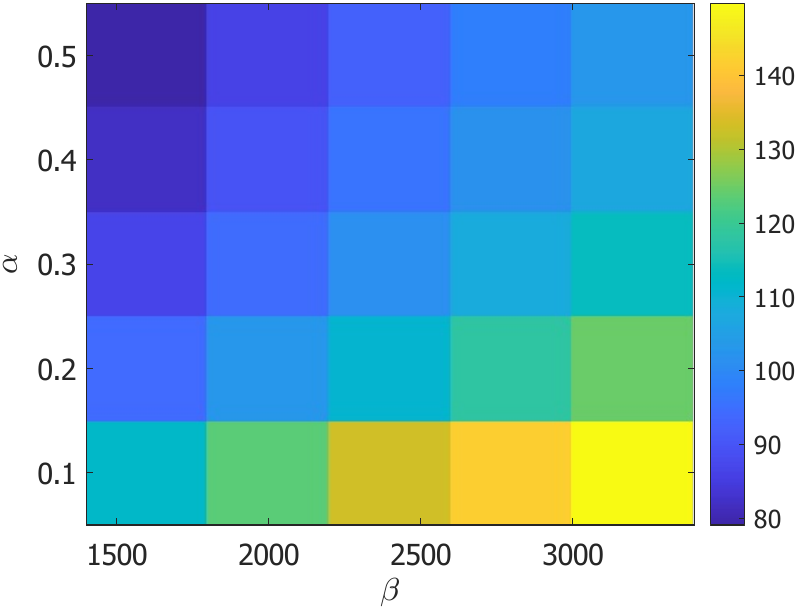}
        \caption{Average $\#S$}
        \label{fig:heatmap-samples}
    \end{subfigure}
    \caption{Heatmaps of average NMSE and number of samples across 100 SoS signals, illustrating the tradeoff between reconstruction accuracy and sampling density as functions of $\{\alpha, \beta\}$.}
    \label{fig:heatmap-comparison}
\end{figure}

Finally, we investigate the effect of the hyperparameters $\alpha$ and $\beta$ on performance.
Fig.~\ref{fig:heatmap-comparison} shows the average NMSE and number of samples across 100 signals for different
parameter choices. Larger values of $\beta$ and smaller values of $\alpha$ generally yield lower
NMSE, but at the cost of increased sampling density.

These results highlight a fundamental tradeoff between reconstruction accuracy and the number of samples. In the next section, we explore whether this trade-off can be further improved through adaptive parameter selection.

\section{Adaptive Selection of $\alpha$ and $\beta$}
\label{sec:adaptive_alpha_beta}

In this section, we present a simple adaptive mechanism for selecting the parameters $\alpha$ and
$\beta$ in the proposed adaptive non-uniform sampling (or VBT-IF-TEM) framework. The objective is
to demonstrate how adaptive parameter selection can further improve sampling efficiency by
aligning the sampling density with the instantaneous behavior of the signal.

As indicated by the inter-sample interval bounds in
\eqref{eq:low_variation_bound_Tn} and \eqref{eq:high_variation_bound_Tn}, the sampling behavior of
the proposed scheme is explicitly governed by the parameters $\alpha$ and $\beta$. Increasing
$\alpha$ relaxes the admissible inter-sample interval and reduces the number of generated
samples, whereas increasing $\beta$ tightens this bound and leads to a higher sampling density.
This intrinsic tradeoff can be exploited to design a more adaptive and robust sampling mechanism
that responds to local signal characteristics, thereby reducing the overall sampling burden
while maintaining reconstruction accuracy.

In many practical signals, large temporal regions exhibit little variation. While the current approach demonstrates strong performance, the use of fixed sampling parameters 
$(\alpha,\beta)$ over the full signal duration results in sampling of low-variation regions, 
contributing to additional redundancy and an increased sample count.

To further enhance performance, we propose a \emph{two-level adaptive sampling strategy} in which
the parameters $(\alpha,\beta)$ are adjusted online based on instantaneous signal activity. We
emphasize that this strategy represents only one possible approach for adapting $\alpha$ and
$\beta$; alternative adaptation rules, switching criteria, or multi-level extensions can be
developed depending on the signal characteristics and application requirements. The present
formulation is chosen for its simplicity and its ability to clearly illustrate the benefits of
adaptive parameter selection.

Specifically, two parameter sets, $(\alpha_1,\beta_1)$ and $(\alpha_2,\beta_2)$, are defined to
correspond to high-rate and low-rate sampling regimes, respectively. The switching mechanism is
driven by monitoring the magnitude of the product $|f(t)\,f'(t)|$ in real time. When this
quantity exceeds a prescribed threshold $\delta$, indicating a region of significant variation,
the sampler operates in the high-rate mode using $(\alpha_1,\beta_1)$, where $\alpha_1$ is small
and $\beta_1$ is large. Conversely, when $|f(t)\,f'(t)| < \delta$, the sampler switches to the
low-rate mode with $(\alpha_2,\beta_2)$, where $\alpha_2$ is larger and $\beta_2$ is smaller,
resulting in substantially reduced sampling density in slowly varying or flat regions.

Importantly, this adaptive modification does not alter the reconstruction algorithm, which
remains identical to the fixed-parameter case. Consequently, the convergence guarantees of the
reconstruction process are preserved, and perfect signal recovery remains possible without
requiring any global upper bound on the inter-sample interval.

We evaluate the proposed adaptive strategy using the chirp bandlimited signal considered
earlier. For these experiments, the parameters are set to $(\alpha_1=0.09,\ \beta_1=2300)$ for
the high-rate mode and $(\alpha_2=0.9,\ \beta_2=10)$ for the low-rate mode, together with
$\delta=6\times10^{-6}$, $c=1$ and $s=4.2c$. The threshold $\delta$ is chosen to be sufficiently small to
ensure that the sampler switches to the low-rate regime only in regions where the signal exhibits
negligible variation.

\begin{figure}[!t]
    \centering
    \includegraphics[width=0.8\linewidth]{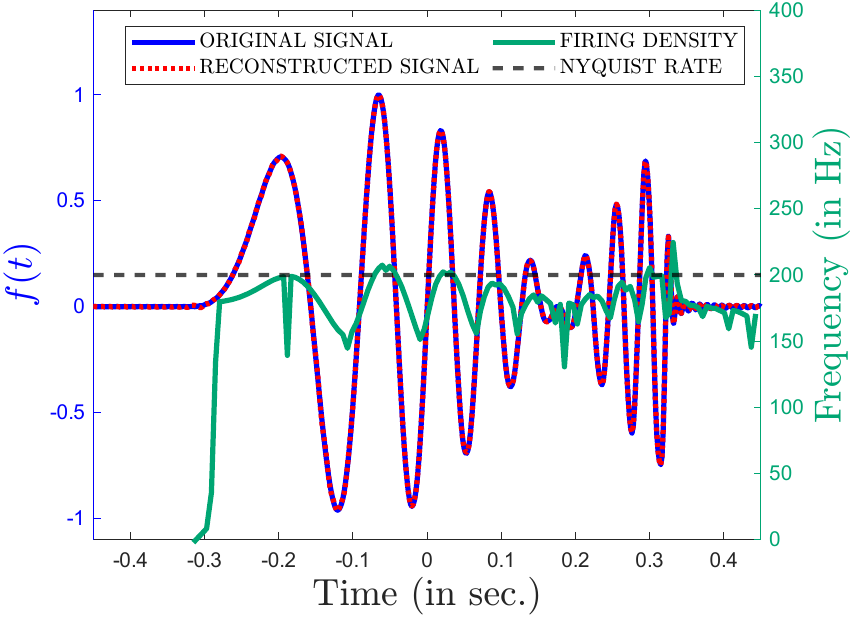}
    \caption{Sampling and reconstruction using Adaptive-NUS with adaptive parameter selection; $\#S=136$; NMSE $=-42.15$ dB.}
    \label{fig:conv-TEM2}
\end{figure}

The reconstructed signal and the corresponding firing density are shown in
Fig.~\ref{fig:conv-TEM2}. As evident from the firing density plot, the sampling rate becomes
extremely low in regions where the signal exhibits little or no variation, while increasing
appropriately during periods of rapid change. Using this adaptive parameter strategy, the number
of samples is reduced to $136$, compared to $169$ samples in the fixed-parameter adaptive scheme
(cf. Fig.~\ref{fig:comparison-table_favo}(c)). For most parts of the signal, the sampling rate
remains below the Nyquist rate, with negligible sampling in flat regions. This reduction in
sampling density comes at the cost of a modest increase in reconstruction error, with the NMSE
increasing from $-54$~dB to $-42.15$~dB, where most of the error contribution arises from flat
signal regions.

Finally, the proposed framework naturally admits further extensions. Additional parameter pairs
$(\alpha_i,\beta_i)$ and alternative switching rules may be incorporated in a systematic manner,
enabling finer adaptation to signal dynamics and allowing the sampling scheme to be tailored to a
broad range of signals and application scenarios.

\section{Practical Applications}
\label{sec:practical_application}

In this section, we demonstrate the applicability of the proposed adaptive non-uniform sampling
framework to practical signal acquisition scenarios. We focus on two representative
applications: ultrasonic guided-wave signals used in structural health monitoring (SHM) and
electrocardiogram (ECG) signals.

\subsection{Ultrasonic Guided-Wave Signals}

Motivated by recent developments in compact guided-wave-based SHM systems, we consider the
ultrasonic pipeline monitoring framework reported in~\cite{Patil2024}. The experimental setup
consists of a $4$-inch-diameter steel pipe instrumented with eight torsional ultrasonic
guided-wave transducers arranged circumferentially around the pipe. Each transducer is mounted
using a $10~\mathrm{mm}\times10~\mathrm{mm}$ ring structure and can operate either as an actuator
or as a sensor.

Damage assessment in such systems is typically performed by comparing the signals received at
the transducers with a baseline signal acquired from the same structure under pristine
conditions. Deviations between the received and baseline signals are interpreted as indicators
of potential structural anomalies.

We apply the proposed adaptive non-uniform sampling methodology directly to the guided-wave
signals recorded by the receiving transducers, with the objective of reducing the sampling
burden while preserving the signal characteristics required for reliable baseline comparison. 
For this experiment, the guided-wave signal has a bandwidth of
$1.2\times10^{5}~\mathrm{Hz}$. The VBT-IF-TEM parameters are set to
$\alpha=0.145$, $\beta=1.999\times10^{10}$, $s=4.2$, and $c=1$.

\begin{figure}[!t]
    \centering
    \includegraphics[width=0.8\linewidth]{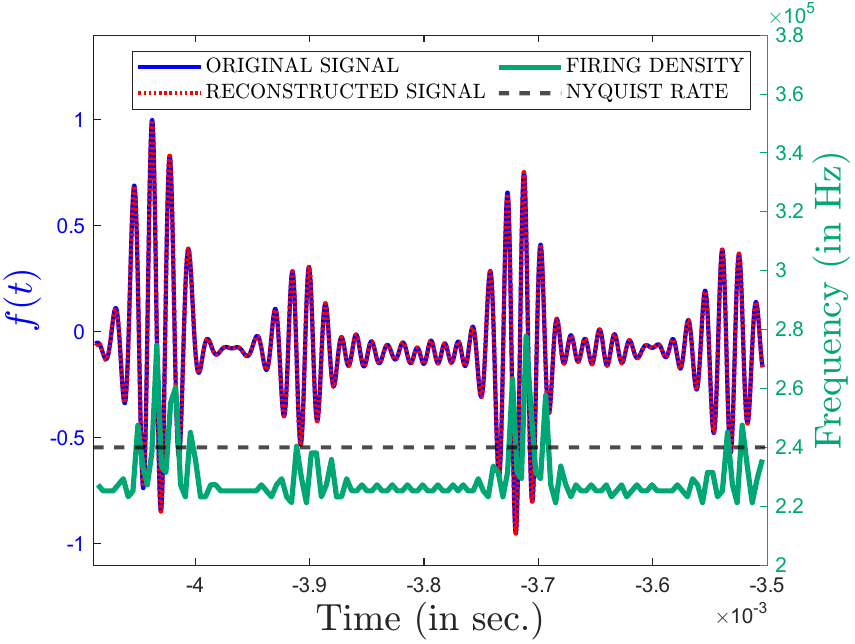}
    \caption{Sampling and reconstruction of an ultrasonic guided-wave signal using the Adaptive-NUS method; $\#S = 1856$;
NMSE = $-42.68\,\mathrm{dB}$. }
    \label{fig:practical1} 
\end{figure}

Fig.~\ref{fig:practical1} illustrates the performance of the proposed scheme for ultrasonic
guided-wave signals. The signal is sampled at a rate well below the Nyquist threshold while still
allowing accurate reconstruction. Specifically, the adaptive sampling process generates $1856$
samples, compared to $1966$ samples required by uniform Nyquist sampling over the same duration.
Despite this reduction in sampling density, the reconstructed signal achieves an NMSE of
$-42.68~\mathrm{dB}$, indicating high reconstruction fidelity.

 These results highlight the
ability of the proposed adaptive sampling framework to exploit local signal structure, enabling
sub-Nyquist sampling while maintaining accurate signal reconstruction in practical SHM
applications.

\subsection{ECG Signals}
We next evaluate the proposed VBT-IF-TEM framework on ECG signals, which are characterized by
strong temporal structure and extended intervals of low activity interspersed with brief,
high-variation events. These properties make ECG signals particularly well-suited for adaptive
sampling strategies.

In this experiment, the ECG signal has a bandwidth of $100~\mathrm{Hz}$, and the VBT-IF-TEM
parameters are set to $\alpha=0.15$, $\beta=5600$, $s=3.2$, and $c=1.0$. These results further
demonstrate the effectiveness of the proposed adaptive framework for biomedical signals, where substantial temporal redundancy can be exploited to achieve efficient sampling without
compromising reconstruction quality.
\begin{figure}[!t]
    \centering
    \includegraphics[width=0.8\linewidth]{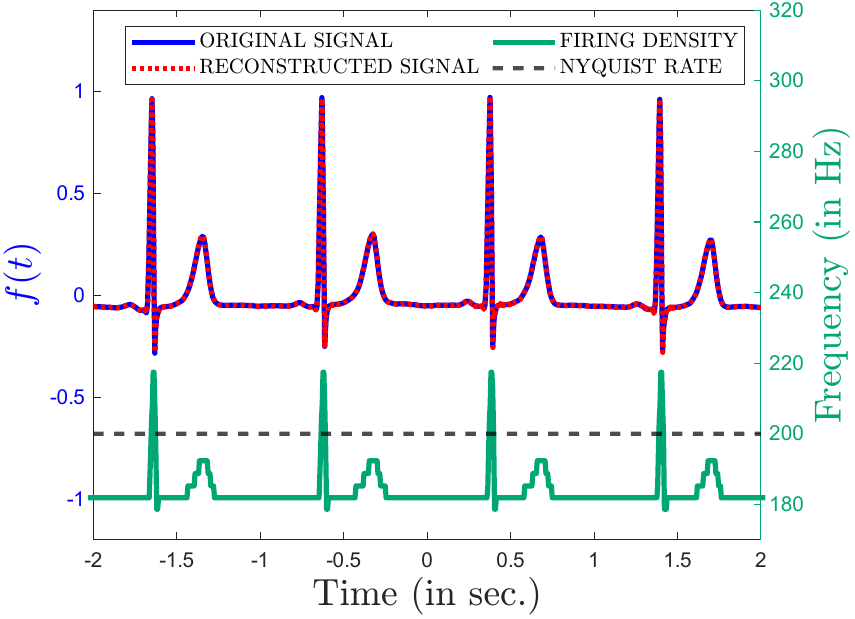}
    \caption{Sampling and reconstruction of an ultrasonic guided-wave signal using the Adaptive-NUS method; $\#S=1837$; NMSE $=-32.45 dB$.}
    \label{fig:practical2}
\end{figure}
Fig.~\ref{fig:practical2} illustrates the application of the proposed scheme to an ECG signal.
Due to the piecewise-smooth nature of the waveform, the adaptive sampler assigns a higher
sampling density to regions of rapid temporal variation, such as the QRS complexes, while
significantly reducing the sampling density in flatter segments. As a result, the proposed
method generates $1837$ samples over the signal duration, compared to $2000$ samples required by
uniform Nyquist sampling. Despite this reduction in sampling load, the reconstructed ECG signal achieves an NMSE of
$-32.45~\mathrm{dB}$, indicating accurate recovery under sub-Nyquist sampling conditions.

\section{Conclusion}
We presented an adaptive non-uniform sampling framework for bandlimited signals based on a locally adaptive sufficient condition for perfect reconstruction. By revisiting the convergence analysis of iterative reconstruction algorithms, we derived a local, energy-based bound that relaxes the need for a global Nyquist-type constraint and permits larger inter-sample intervals in slowly varying signal regions.

Guided by this analysis, we developed a variable-bias, variable-threshold integrate-and-fire time encoding machine that generates signal-dependent time encodings while preserving
reconstruction guarantees. A biased-signal formulation was introduced to suppress excessive firing in low-energy regimes where
the magnitude of the signal amplitude is small. Simulations and real-world experiments demonstrated that the proposed framework reduces sampling density compared to uniform sampling and conventional
IF-TEMs while maintaining accurate reconstruction, and an adaptive parameter selection strategy illustrated how the sampling–accuracy tradeoff can be navigated in practice.

\bibliographystyle{elsarticle-num} 
\bibliography{references,refs21,refs,tem_refs,tem_ref2,refs2,ref_tem_AA_YK}

@INPROCEEDINGS{diff_and_fire,
  author={Kamath, Abijith Jagannath and Seelamantula, Chandra Sekhar},
  booktitle={Proc. IEEE Int. Conf. Acoust., Speech and Signal Process. (ICASSP)}, 
  title={Differentiate-and-Fire Time-Encoding of Finite-Rate-of-Innovation Signals}, 
  year={2022},
  volume={},
  number={},
  pages={5637-5641},
  doi={10.1109/ICASSP43922.2022.9746159}}

@article{thao2020time,
  title={Time encoding of bandlimited signals: Reconstruction by pseudo-inversion and time-varying multiplierless {FIR} filtering},
  author={Thao, Nguyen T and Rzepka, Dominik},
  journal={IEEE Trans. Signal Process.},
  volume={69},
  pages={341--356},
  year={2020},
  publisher={IEEE}
}

@article{florescu2025generalized,
  title={A generalized approach for recovering time encoded signals with finite rate of innovation},
  author={Florescu, Dorian},
  journal={IEEE Trans. Signal Process.},
  year={2025},
  publisher={IEEE}
}

@article{florescu2022time,
  title={Time encoding via unlimited sampling: Theory, algorithms and hardware validation},
  author={Florescu, Dorian and Bhandari, Ayush},
  journal={IEEE Trans. Signal Process.},
  volume={70},
  pages={4912--4924},
  year={2022},
  publisher={IEEE}
}

@INPROCEEDINGS{AdaptiveBiasIFTEM,
  author={Omar, Aseel and Cohen, Alejandro},
  booktitle={Proc. European Signal Process. Conf. (EUSIPCO)}, 
  title={Adaptive Integrate-and-Fire Time Encoding Machine}, 
  year={2024},
  volume={},
  number={},
  pages={2442-2446},
  doi={10.23919/EUSIPCO63174.2024.10715196}}

@inproceedings{hila2022time,
  title={Time-based quantization for {FRI} and bandlimited signals},
  author={Naaman, Hila and Mulleti, Satish and Eldar, Yonina C and Cohen, Alejandro},
  booktitle={European Signal Process. Conf. (EUSIPCO)},
  pages={2241--2245},
  year={2022}
}

@article{naaman2025time,
  title={Time Encoding Quantization of Bandlimited and Finite-Rate-of-Innovation Signals},
  author={Naaman, Hila and Bernardo, Neil Irwin and Cohen, Alejandro and Eldar, Yonina C},
  journal={IEEE Access},
  year={2025},
  publisher={IEEE}
}

@article{bernardo2025symbol,
  title={Symbol Detection Using an Integrate-and-Fire Time Encoding Receiver},
  author={Bernardo, Neil Irwin},
  journal={arXiv preprint arXiv:2508.17704},
  year={2025}
}

@ARTICLE{differentTEMS2,
  author={Naaman, Hila and Mulleti, Satish and Eldar, Yonina C.},
  journal={IEEE Trans. Signal Process.}, 
  title={{FRI-TEM: Time Encoding Sampling of Finite-Rate-of-Innovation Signals}}, 
  year={2022},
  volume={70},
  number={},
  pages={2267-2279},
  doi={10.1109/TSP.2022.3167146}}

@ARTICLE{differentTEMS3,
  author={Alexandru, Roxana and Dragotti, Pier Luigi},
  journal={IEEE Trans. Signal Process.}, 
  title={Reconstructing Classes of Non-Bandlimited Signals From Time Encoded Information}, 
  year={2020},
  volume={68},
  number={},
  pages={747-763},
  doi={10.1109/TSP.2019.2961301}}

@inproceedings{kamath2023multichannel,
  title={Multichannel time-encoding of finite-rate-of-innovation signals},
  author={Kamath, Abijith Jagannath and Seelamantula, Chandra Sekhar},
  booktitle={Proc. IEEE Int. Conf. Acoust., Speech and Signal Process. (ICASSP)},
  pages={1--5},
  year={2023},
  organization={IEEE}
}

@article{fu2024time,
  title={Time-based finite-rate-of-innovation sampling for variable-pulse-width signal},
  author={Fu, Ning and Zhang, Hongyi and Yun, Shuangxing and Wei, Zhiliang and Qiao, Liyan},
  journal={IEEE Transa. Inst. Meas.},
  volume={73},
  pages={1--9},
  year={2024},
  publisher={IEEE}
}

@article{Patil2024,
author={Patil, Sheetal and Banerjee, Sauvik and Tallur, Siddharth},
title={Smart structural health monitoring (SHM) system for on-board localization of defects in pipes using torsional ultrasonic guided waves},
journal={Scientific Reports},
volume={14},
number={1},
pages={24455},
year={2024},
doi={10.1038/s41598-024-76236-w}
}

@inproceedings{andreolli2023sampling,
  title={Sampling theorems in spaces of variable bandwidth generated via Wilson basis},
  author={Andreolli, Beatrice and Gr{\"o}chenig, Karlheinz},
  booktitle={Proc. Intl. Conf. Sampling theory and Appl. (SampTA)},
  pages={1--5},
  year={2023},
  organization={IEEE}
}

@article{clark1985transformation,
  title={A transformation method for the reconstruction of functions from nonuniformly spaced samples},
  author={Clark, J and Palmer, M and Lawrence, PO},
  journal={IEEE Tran. Acoustics, Speech, Signal Process.},
  volume={33},
  number={5},
  pages={1151--1165},
  year={1985},
  publisher={IEEE}
}

@article{dvorkind2008nonlinear,
  title={Nonlinear and nonideal sampling: {T}heory and methods},
  author={Dvorkind, Tsvi G and Eldar, Yonina C and Matusiak, Ewa},
  journal={IEEE Trans. Signal Process.},
  volume={56},
  number={12},
  pages={5874--5890},
  year={2008},
  publisher={IEEE}
}

@inproceedings{martinez2016amplitude,
  title={Amplitude sampling},
  author={Mart{\'\i}nez-Nuevo, Pablo and Lai, Hsin-Yu and Oppenheim, Alan V},
  booktitle={Allerton Conf. Comm. Ctr. Comput. (Allerton)},
  pages={17--22},
  year={2016},
  organization={IEEE}
}

@article{horiuchi1968sampling,
  title={Sampling principle for continuous signals with time-varying bands},
  author={Horiuchi, Kazuo},
  journal={Info. Control},
  volume={13},
  number={1},
  pages={53--61},
  year={1968},
  publisher={Elsevier}
}

@inproceedings{wei2007sampling,
  title={Sampling based on local bandwidth},
  author={Wei, Dennis and Oppenheim, Alan V},
  booktitle={Asilomar Conf. Signals, Syst. Comput},
  pages={1103--1107},
  year={2007},
  organization={IEEE}
}

@inproceedings{rzepka2013recovery,
  title={Recovery of varying-bandwidth signal from samples of its extrema},
  author={Rzepka, Dominik and Mi{\'s}kowicz, Marek},
  booktitle={Signal Processing: Algo. Architectures Arrangements, Appl.(SPA)},
  pages={143--148},
  year={2013},
  organization={IEEE}
}

@article{boche2012towards,
  title={Towards a general theory of reconstruction of bandlimited signals from sine wave crossings},
  author={Boche, Holger and M{\"o}nich, Ullrich J},
  journal={Signal Process.},
  volume={92},
  number={3},
  pages={737--751},
  year={2012},
  publisher={Elsevier}
}

@article{li2006remarks,
  title={Remarks on the {V}oronoi method in {P}aley--{W}iener space},
  author={Li, Song-Hua and Lin, Wei},
  journal={J. Mathematical Ana. Appl.},
  volume={318},
  number={1},
  pages={1--14},
  year={2006},
  publisher={Elsevier}
}

@article{sun2002reconstruction,
  title={Reconstruction of band-limited signals from local averages},
  author={Sun, Wenchang and Zhou, Xingwei},
  journal={IEEE Trans. Info. Theory},
  volume={48},
  number={11},
  pages={2955--2963},
  year={2002},
  publisher={IEEE}
}

@ARTICLE{song2012_avgSampling,
  author={Song, Zhanjie and Liu, Bei and Pang, Yanwei and Hou, Chunping and Li, Xuelong},
  journal={IEEE Trans. Info. Theory}, 
  title={An Improved {N}yquist–{S}hannon Irregular Sampling Theorem From Local Averages}, 
  year={2012},
  volume={58},
  number={9},
  pages={6093-6100},
  doi={10.1109/TIT.2012.2199959}}

@ARTICLE{maravasti1991,
  author={Marvasti, F. and Analoui, M. and Gamshadzahi, M.},
  journal={IEEE Trans. Signal Process.}, 
  title={Recovery of signals from nonuniform samples using iterative methods}, 
  year={1991},
  volume={39},
  number={4},
  pages={872-878},
  doi={10.1109/78.80909}}

@article{wiley1978recovery,
  title={Recovery of bandlimited signals from unequally spaced samples},
  author={Wiley, R},
  journal={IEEE Trans. Comm.},
  volume={26},
  number={1},
  pages={135--137},
  year={1978},
  publisher={IEEE}
}

@article{duffin1952class,
  title={A class of nonharmonic {F}ourier series},
  author={Duffin, Richard J and Schaeffer, Albert C},
  journal={Trans. American Math. Soc.},
  volume={72},
  number={2},
  pages={341--366},
  year={1952},
  publisher={JSTOR}
}

@article{grochenig1992reconstruction,
  title={Reconstruction algorithms in irregular sampling},
  author={Gr{\"o}chenig, Karlheinz},
  journal={Mathematics of computation},
  volume={59},
  number={199},
  pages={181--194},
  year={1992}
}

@incollection{marvasti1993nonuniform,
  title={Nonuniform sampling},
  author={Marvasti, Farokh},
  booktitle={Advanced topics in Shannon sampling and interpolation theory},
  pages={121--156},
  year={1993},
  publisher={Springer}
}

@incollection{feichtinger2021theory,
  title={Theory and practice of irregular sampling},
  author={Feichtinger, Hans G and Gr{\"o}chenig, Karlheinz},
  booktitle={Wavelets},
  pages={305--363},
  year={2021},
  publisher={CRC Press}
}

@article{Liang1999,
  author = {Liang, Zhi-Pei and Lauterbur, Paul C.},
  title = {Principles of Magnetic Resonance Imaging: {A} Signal Processing Perspective},
  journal = {IEEE Press Series in Biomedical Engineering},
  publisher = {IEEE Press},
  year = {1999},
  note = {Chapter on non-uniform sampling in NMR/MRI}
}

@article{Beatty2005,
  author = {Beatty, Patrick J. and Nishimura, Dwight G. and Pauly, John M.},
  title = {Rapid Gridding Reconstruction With a Recursive, Spatially Oversampled Fourier Transform},
  journal = {IEEE Trans. Medical Imaging},
  volume = {24},
  number = {1},
  pages = {16--25},
  year = {2005},
  doi = {10.1109/TMI.2004.837765}
}

@article{shannon,
	Author = {Shannon, C. E.},
	Journal = {The Bell Syst. Tech. J.},
	Pages = {623-656},
	Title = {A Mathematical Theory of Communication},
	Volume = 27,
	Year = 1948}

@ARTICLE{nyquist,
author={Nyquist, H.},
journal={Trans. American Inst. of Elect. Eng.},
title={Certain Topics in Telegraph Transmission Theory},
year={1928},
volume={47},
number={2},
pages={617-644},
doi={10.1109/T-AIEE.1928.5055024},
ISSN={0096-3860},
month={Apr.},}

@article{landau,
	Author = {Landau, H. J.},
	Doi = {10.1007/BF02395039},
	Issn = {0001-5962},
	Journal = {Acta Mathematica},
	Language = {English},
	Number = {1},
	Pages = {37-52},
	Publisher = {Kluwer Academic Publishers},
	Title = {Necessary density conditions for sampling and interpolation of certain entire functions},
	Volume = {117},
	Year = {1967}}

@article{malmirchegini2015non,
  title={Non-uniform sampling based on an adaptive level-crossing scheme},
  author={Malmirchegini, Mehrzad and Kafashan, Mohammad Mehdi and Ghassemian, Mona and Marvasti, Farokh},
  journal={IET Signal Process.},
  volume={9},
  number={6},
  pages={484--490},
  year={2015},
  publisher={Wiley Online Library}
}

@article{pandey2024cardinality,
  title={Cardinality constraint non-uniform sampling for maximizing reconstruction accuracy of time-varying signals},
  author={Pandey, Sakshi and Banerjee, Amit},
  journal={IEEE Trans. Info. Theory},
  volume={70},
  number={9},
  pages={6746--6756},
  year={2024},
  publisher={IEEE}
}

@article{dar2019high,
  title={On high-resolution adaptive sampling of deterministic signals},
  author={Dar, Yehuda and Bruckstein, Alfred M},
  journal={J. Mathematical Imaging and Vision},
  volume={61},
  number={7},
  pages={944--966},
  year={2019},
  publisher={Springer}
}

@article{feizi2014backward,
  title={Backward adaptation for power efficient sampling},
  author={Feizi, Soheil and Angelopoulos, Georgios and Goyal, Vivek K and M{\'e}dard, Muriel},
  journal={IEEE Trans. Signal Process.},
  volume={62},
  number={16},
  pages={4327--4338},
  year={2014},
  publisher={IEEE}
}

@article{feizi2012time,
  title={Time-stampless adaptive nonuniform sampling for stochastic signals},
  author={Feizi, Soheil and Goyal, Vivek K and M{\'e}dard, Muriel},
  journal={IEEE Trans. Signal Process.},
  volume={60},
  number={10},
  pages={5440--5450},
  year={2012},
  publisher={IEEE}
}

@article{lazar2004time,
  title={Time encoding with an integrate-and-fire neuron with a refractory period},
  author={Lazar, Aurel A.},
  journal={Neurocomputing},
  volume={58},
  pages={53--58},
  year={2004},
  publisher={Elsevier}
}

@article{logan1977information,
  title={Information in the zero crossings of bandpass signals},
  author={Logan Jr, Benjamin F},
  journal={Bell Syst. Technical J.},
  volume={56},
  number={4},
  pages={487--510},
  year={1977},
  publisher={Wiley Online Library}
}

@article{gontier2014sampling,
  title={Sampling based on timing: Time encoding machines on shift-invariant subspaces},
  author={Gontier, David and Vetterli, Martin},
  journal={Applied and Comput. Harmonic Anal.},
  volume={36},
  number={1},
  pages={63--78},
  year={2014},
  publisher={Elsevier}
}

@ARTICLE{adam2020encoding,
  author={Adam, Karen and Scholefield, Adam and Vetterli, Martin},
  journal={IEEE Trans. Signal Process.}, 
  title={Sampling and Reconstruction of Bandlimited Signals With Multi-Channel Time Encoding}, 
  year={2020},
  volume={68},
  number={},
  pages={1105-1119},
  doi={10.1109/TSP.2020.2967182}}

@ARTICLE{yao_nonuniform,  author={Kung Yao and Thomas, J.},  journal={IEEE Trans. Circuit Theory},   title={On Some Stability and Interpolatory Properties of Nonuniform Sampling Expansions},   year={1967},  volume={14},  number={4},  pages={404-408},  doi={10.1109/TCT.1967.1082745}}

@article{rzepka2018reconstruction,
  title={Reconstruction of signals from level-crossing samples using implicit information},
  author={Rzepka, Dominik and Mi{\'s}kowicz, Marek and Ko{\'s}cielnik, Dariusz and Thao, Nguyen T},
  journal={IEEE Access},
  volume={6},
  pages={35001--35011},
  year={2018},
  publisher={IEEE}
}

@article{mark1981nonuniform,
  title={A nonuniform sampling approach to data compression},
  author={Mark, J and Todd, T},
  journal={IEEE Trans. Comm.},
  volume={29},
  number={1},
  pages={24--32},
  year={1981},
  publisher={IEEE}
}

@article{sayiner1996level,
  title={A level-crossing sampling scheme for {A/D} conversion},
  author={Sayiner, Necip and Sorensen, Henrik V and Viswanathan, Thayamkulangara R},
  journal={IEEE Trans. Circuits Sys. II: Analog Digital Signal Process.},
  volume={43},
  number={4},
  pages={335--339},
  year={1996},
  publisher={IEEE}
}

@ARTICLE{tsivids2010tutorial,
  author={Tsividis, Yannis},
  journal={IEEE Trans. Circuits Syst. II, Exp. Briefs}, 
  title={Event-Driven Data Acquisition and Digital Signal Processing— {A} Tutorial}, 
  year={2010},
  volume={57},
  number={8},
  pages={577-581},
  doi={10.1109/TCSII.2010.2056012}}

@INPROCEEDINGS{VBIFTEM,
  author={Arora, Anshu and Mulleti, Satish},
  booktitle={IEEE Int. Conf. Acoust., Speech and Signal Process. (ICASSP)}, 
  title={A Lowrate Variable-Bias Integrate-and-Fire Time Encoding Machine}, 
  year={2025},
  volume={},
  number={},
  pages={1-5},
  doi={10.1109/ICASSP49660.2025.10888540}}

@inproceedings{omar2024adaptive,
  title={Adaptive integrate-and-fire time encoding machine},
  author={Omar, Aseel and Cohen, Alejandro},
  booktitle={European Signal Process. Conf. (EUSIPCO)},
  pages={2442--2446},
  year={2024},
  organization={IEEE}
}

@article{lazar2004perfect,
  title={Perfect recovery and sensitivity analysis of time encoded bandlimited signals},
  author={Lazar, Aurel A and T{\'o}th, L{\'a}szl{\'o} T},
  journal={IEEE Trans Circuits and Syst. I: Regular Papers},
  volume={51},
  number={10},
  pages={2060--2073},
  year={2004},
  publisher={IEEE}
}

@article{feichtinger1995efficient,
  title={Efficient numerical methods in non-uniform sampling theory},
  author={Feichtinger, Hans G and Gr\"ochenig, Karlheinz and Strohmer, Thomas},
  journal={Nume. Math.},
  volume={69},
  pages={423--440},
  year={1995},
  publisher={Springer}
}

@ARTICLE{yen_nonuniform,  author={Yen, J.},  journal={IRE Trans. Circuit Theory},   title={On Nonuniform Sampling of Bandwidth-Limited Signals},   year={1956},  volume={3},  number={4},  pages={251-257},  doi={10.1109/TCT.1956.1086325}}

\end{document}